\documentclass{article}
 \usepackage{color}
\usepackage{amsfonts}
\usepackage{amsmath}
\usepackage{amssymb}
\usepackage{tikz}
\usepackage{graphicx}
\usepackage{comment}

\newcommand{\ba}{\begin{eqnarray}}
\newcommand{\ea}{\end{eqnarray}}
\newcommand{\nn}{\nonumber}
\newcommand{\cW}{{\mathcal{W}}}

\newcommand{\cc}{\mathbf{c}}
\newcommand{\Z}{{\mathcal{Z}}}

\newcommand{\be}{\begin{equation}}
\newcommand{\ee}{\end{equation}}

\newcommand{\cL}{\mathcal{L}}

\def\a{\alpha}
\def\b{\beta}

\def\d{\delta}
\def\e{\epsilon}

\def\k{\kappa}
\def\l{\lambda}
\def\m{\mu}
\def\n{\nu}

\def\w{\omega}

\def\D{\Delta}

\def\L{\Lambda}

\makeatletter

    \@addtoreset{equation}{section}
\makeatother

\def\Zv{\mathcal{Z}_\text{vect}}
\def\Zbf{\mathcal{Z}_\text{bif}}

\def\bZbf{\bar{\mathcal{Z}}_\text{bif}}

\def\aY{|\vec{a},\vec{Y}\rangle}

\newcommand{\superp}[2]{\genfrac{}{}{0pt}{}{#1}{#2}}

\allowdisplaybreaks[1]

\begin{document}
%
%
%
%
%

\begin{center}
{\Large Nekrasov and Argyres-Douglas theories \\ 
in  spherical Hecke algebra representation}
\vskip 1cm
{\large Chaiho Rim  and Hong Zhang\footnote{Address after Nov 2016: Institute of Theoretical Physics, Chinese Academy of Sciences, Beijing 100190, P. R. China}}
\vskip 5mm
{\it Department of Physics and Center for Quantum Spacetime (CQUeST)}\\
{\it Sogang University, Seoul 121-742, Korea}
\end{center}

\vskip 10mm

\begin{abstract}
AGT conjecture connects  Nekrasov instanton partition function of 4D quiver gauge theory with  2D Liouville conformal blocks. 
We re-investigate this connection using the
central extension of spherical Hecke algebra 
in  q-coordinate representation,
q being the instanton expansion parameter.
Based on AFLT basis together with interwiners
we construct  gauge conformal state 
and demonstrate its equivalence to the Liouville conformal state, 
with careful attention to  the proper scaling behavior of the state.  
Using the colliding limit of regular states, 
we obtain the formal expression 
of irregular conformal states corresponding to Argyres-Douglas theory, 
which involves summation of functions over Young diagrams.
\end{abstract}

\vskip 12mm

\setcounter{footnote}{0}
\section{Introduction}
Liouville conformal block is a useful tool to understand 
$SU(2)$ Nekrasov partition function of 4D quiver gauge theory 
due to AGT conjecture \cite{AGT_2009} 
and is generalized to Toda theory \cite{Wyllard_2009}
which represents $SU(N)$ Nekrasov partition function. 
The Virasoro  conformal state 
is soon generalized in \cite{Gaiotto_2009} 
where a new conformal state is constructed. 
The new state is related with  asymptotically free SU(2) 
quiver gauge theories, 
which reproduce irregular singularities of the Seiberg-Witten curve 
corresponding to the Argyres-Douglas theory \cite{AD,APSW}.
The new state is a kind of coherent (rather than primary) state 
and is called Gaiotto state in the physics community. 
Among mathematicians, however, 
the state is  known as Whitaker state 
\cite{Whittaker} in earlier stage.  
We will call the new state ``irregular conformal state",
and the conformal state corresponding to the  Nekrasov partition function
``regular conformal state". 

The irregular state is of interest 
because the irregular conformal block is given 
as the inner product of two irregular states. 
For example,
two states $|\Delta,\Lambda^2 \rangle$ and $|\Delta,\Lambda,m\rangle$ 
provide such inner products:
$ \langle \Delta,\Lambda^2|\Delta,\Lambda^2 \rangle$
produces the partition function of 
$SU(2)$ with $N_f = 0$, 
and 
$ \langle \Delta,\Lambda^2|\Delta,\Lambda,m \rangle$ 
produces that of $SU(2)$ with $N_f = 1$. 
The special feature of these 
irregular states is that, 
they have non-vanishing eigenvalues 
under the action of certain Virasoro positive modes. 
For example, the states above considered 
have the property for $k\geq 1$,
$L_k |\Delta,\Lambda^2\rangle
= \delta_{k,1}\Lambda^2|\Delta,\Lambda^2 \rangle$   
and 
$L_k |\Delta,\Lambda,m\rangle
= (\delta_{k,1} m\Lambda
+\delta_{k,2}  \Lambda^2) 
|\Delta,\Lambda,m\rangle $.

Systematic construction of the irregular state is 
first given in terms of a limiting process in  \cite{MMM2009}.
Four point conformal block provides the irregular state 
$
|\Delta,\Lambda^2\rangle = \sum_Y
\Lambda^{2|Y|}
Q_\Delta^{-1}\Big([1^{|Y|}],\ Y\Big) L_{-Y}|\Delta\rangle
$
using the Shapovalov form 
$Q_\Delta \Big(Y,\ Y'\Big) 
= \langle \Delta | L_{Y'} L_{-Y} |\Delta \rangle $ , 
where  $L_{-Y} |\Delta \rangle  =L_{-k_d}\dots L_{-k_2}L_{-k_1}|\Delta \rangle$ represents 
descendant with proper ordering of the Young diagram
 $Y=\{k_1\geq k_2\geq \dots \geq k_d>0\}$.  

Likewise, 
$
|\Delta,\Lambda,m\rangle = \sum_Y \sum_{p}
m^{|Y|-2p}\Lambda^{|Y|}
Q_\Delta^{-1}\Big([2^p,1^{|Y|-2p}],\ Y\Big) L_{-Y}|\Delta\rangle
$.
This representation is generalized 
into the simultaneous eigenvector 
of two generators $L_1$ and $L_n$  in \cite{BMT_2011}.
In addition, Virasoro irregular state of 
higher rank $n$ 
(simultaneous eigenstate 
of $ L_k$ with  $n \leq k \leq 2n $) 
is suggested in \cite{KMST2013},
while some of  the coefficients for the  representation 
are not fixed. 

Colliding limit,  a limiting procedure  to obtain the
 irregular state  from the regular state 
is clarified in \cite{GT2012}.  
The decoupling limit  in  \cite{MMM2009} 
and also in the  matrix model  \cite{EM_2009}  
is a special case of the colliding limit. 
The colliding limit turns out to be a very efficient tool to investigate 
the irregular state to find 
the correct representation of the irregular state 
of rank greater than 1. 
Indeed, the coefficients 
undetermined in  \cite{KMST2013} 
are fixed by irregular matrix model in \cite{CRZ_2015}  
which obeys  consistency conditions 
of Virasoro generators of lower positive modes
$L_k$ with $0 \le k <n$.  
The irregular matrix model analysis is extended to $\cW$-symmetry  in \cite{CRZ_2016}.  

The irregular matrix model analysis, however, 
provides indirect information  
because the partition function of the matrix model 
is equivalent to the  inner-product
of two states. Direct process 
to find the irregular state is more desirable. 
For this goal, we resort to the representation 
of spherical double degenerate affine Hecke algebra 
(spherical DDAHA or  SH  for short). 
DDAHA is generated by $z_i$ 
and ${\cal D}_i = z_i \nabla_i +\sum_{j<i} \sigma_{ij} $ 
($i=1, \cdots, N$) where $\nabla_i$ is the 
Dunkl operator and $\sigma_{ij}$ is the transposition 
of $z_i$ to $z_j$. 
Spherical DDAHA (SH) is restricted to the symmetric part 
of product of $ z_i$'s and ${\cal D}_i $'s. 
SH also allows central extension, 
which is considered in this text and is still denoted 
as SH instead of SH$^c$  for simplicity.  
More details refer to \cite{r:SV}.
The algebraic elements and their  commutation relations 
are given in section 2. 

In this paper, we elaborate and generalize
the procedure presented in \cite{MRZ_2014}
to irregular Virasoro state of arbitrary rank $m$
using SH algebra 
based on AFLT orthonormal basis \cite{Alba2011}
and interwiners \cite{MO_2012, BMZ_2015} .
For this purpose we construct 
the regular conformal state  $|T_m\rangle$ 
in $q$-coordinates ($q$ being the instanton expansion parameter), 
which is the counter-part of Liouville conformal state 
$|R_m\rangle$  used in \cite{GT2012}. 
The equivalence relation is  manifest 
after the proper scaling behavior is compensated. 
After this, one can find the irregular state  
 $|I_m \rangle$ of rank $m$
using the colliding limit. 

This paper is organized as follows: 
In section 2  we briefly introduce the spherical Hecke algebra,
AFLT basis and the interwiner. 
Based on these elements, we construct 
the $q$-representation of 
the gauge conformal state, 
counter-part of
holomorphic representation of 
Liouville conformal state.
In section 3, we investigate 
the $q$-representation of 
the Heisenberg and Virasoro representation
using the SH algebra, 
and find the equivalence relation of 
the gauge conformal state with 
the Liouville conformal state.
The equivalence is established according to AGT dictionary 
as the consequence of the proper scaling of the $q$-basis.  
In section 4, we find the formal solution of 
irregular state using the colliding limit. 
Section 5 is the conclusion and 
details of Hecke algebra calculations are collected in the appendix.

\section{Spherical Hecke algebra and its representation}
In this section, we  construct regular states 
using the  spherical Hecke  algebra with central extension, 
based on the AFLT basis with interwiners.

\subsection{Spherical Hecke generators}
We summarize the property of Spherical Hecke algebra,
the details of which can be found  in  \cite{Kanno:2013aha}. 
The SH algebra has generators $D_{r,s}$  with $r$ integer and 
$s$ non-negative integer.  
The first index $r$  is called degree 
and the second one  $s$  order of generator.

The commutation relations between  generators  of degree $\pm 1, 0$ 
are the defining relations  \cite{r:SV},
\ba
\left[D_{0,l} , D_{1,k} \right] & =& D_{1,l+k-1}, \;\;\; l \geq 1 \,,
\label{SH1} \\
\left[D_{0,l},D_{-1,k}\right]&=&-D_{-1,l+k-1}, \;\;\; l \geq 1 \,,
\label{SH2}\\
\left[D_{-1,k},D_{1,l}\right]&=&E_{k+l} \;\;\; l,k \geq 0\label{eDDE}\,,\\
\left[D_{0,l} , D_{0,k} \right] & =& 0 \,,\,\, k,l\geq 0\,,
\label{SH4}
\ea
where $E_k$ is a nonlinear combination of $D_{0,k}$,
determined by a generating function,{\footnote
{we follow the notation in \cite{BMZ_2015} where the omega  background
parameters $\e_1,\e_2$ are used instead of the CFT parameter 
$\b=-\e_1/\e_2$ in \cite{MRZ_2014}. 
The comparison between the two are given as:
$ D_{0,n+1}=(-\e_2)^n \tilde D_{0,n+1},\quad D_{\pm 1,n}=(-\e_2)^n \tilde D_{\pm 1,n},\quad
E_{n}=(-\e_2)^n \tilde E_n\,,
$ and $\cc_{n}=(-\e_2)^n \tilde \cc_n$.
Tilde is used for the ones in  \cite{MRZ_2014}.}
\be
1-\e_+\sum_{l\geq 0}E_l s^{l+1}= \exp(\sum_{l\geq 0}(-1)^{l+1}\cc_l \pi_l(s))\exp(\sum_{l\geq 0}D_{0,l+1} \w_l(s)).
\label{com0}
\ee
Here
$\pi_l(s)=s^l G_l(1-\e_+s)$ 
and $\w_l(s)=\sum_{q=-\e_1,-\e_2,\e_+}s^l(G_l(1-qs)-G_l(1+qs))$.
We use  notations  
$G_0(s)=-\log(s)$,  $G_l(s)=(s^{-l}-1)/l $ for $ l \geq 1$ and  $\e_+=\e_1+\e_2$.
$\cc_l$ ($l\geq 0$) is the  central extension and 
plays an  essential  role in  comparing  with the conformal algebra. 
Some of the explicit expressions of $E_\ell$ are given as follows;  
\begin{align}
E_0&=\mathbf{c}_0,
\nn\\
E_1&=-\mathbf{c}_1 -\cc_0(\cc_0-1)\e_+ /2,\label{E_1} 
\nn\\
E_2&=\cc_2-\cc_1(1-\cc_0)\e_+ +\cc_0(\cc_0-1)(\cc_0-2)\e_+^2 /6 -2\e_1\e_2 D_{0,1}.
\end{align}
Other generators $D_{\pm r, l}$ for $l\geq 0, r>1$ 
are defined recursively as:
\begin{align}
D_{l+1,0} &= \frac1l \left[D_{1,1} , D_{l,0} \right] ,~~~
D_{-l-1,0} = \frac1l \left[D_{-l,0},D_{-1,1}\right] \,,
\nn \\
D_{r,l} &= \left[D_{0,l+1} , D_{r,0} \right] ,  ~~~~~~
D_{-r,l}= \left[D_{-r,0} , D_{0,l+1} \right]\,. 
\end{align}

It is noted that SH contains the Heisenberg   
and Virasoro algebras whose generators 
are identified as \cite{r:SV},
\begin{align} 
 J_{n}&=(-\sqrt{-\e_1\e_2})^{-|n|} D_{-n,0}  ~~~{\rm for} \; n\neq 0,
\label{defJ}\\
 L_n&=(-\sqrt{-\e_1\e_2})^{-|n|} D_{-n,1}/|n| -(1-|n|) \cc_0\, \e_+ J_n/2
 ~~~{\rm for} \; n\neq 0
\label{defV}
\end{align}  
Zero mode $J_0$ is defined using $E_1$ \eqref{E_1},
\be 
J_0=E_1/(-\e_1\e_2) ,
\label{def:J_0}
\ee
and $L_0$ is derived from $L_0=[L_1,L_{-1}]/2$,
\be
L_0=  E_2/(-2\e_1\e_2).
\label{def:L_0}
\ee   

The commutation relations among these Heisenberg   
and Virasoro generators are,
\begin{align}
\left[ J_n, J_m\right] &= \frac{n \cc_0}{\beta} \delta_{n+m,0},\\
\left[L_n, J_m\right] &= -m J_{n+m},\\
\left[ L_n, L_m\right]&= (n-m) L_{n+m}+\frac{c}{12}(n^3-n) \delta_{n+m,0}\,,
\label{Hecke-CFT}
\end{align}
with the central charge $c=  \left(
\cc_0\e_2^2 -\cc_0\e_2 \e_+ + \cc_0 \e_+^2  - \cc_0^3 \e_+^2 
 \right)/(-\e_1\e_2)$. 
\subsection{Gauge conformal state for Nekrasov partition function}
The Nekrasov partition function of  $ U(N)^{\otimes n}$ linear quiver gauge theory on  a Riemann sphere 
is given in terms of $n+3$ punctures, 
and its instanton part is given by
\ba
Z_{\text {inst}}^{(n+3)-point}(q_1, \dots, q_n)
=&&\sum_{\vec Y_1,\cdots, \vec Y_n}
\prod_{i=1}^n q_i^{|\vec Y_i|}\Zv (\vec a^{(i)}, \vec Y_i) \prod_{i=1}^{n-1}  \Zbf(\vec a^{(i)},\vec Y_i;\vec a^{(i+1)},\vec Y_{i+1}|\n_{i}) \nn\\
&& \times \prod_{I=1}^N \Z_{\text {fund}}(\vec a^{(1)}, \vec Y_1,\tilde \m_I )
\Z_{\text {afd}}(\vec a^{(n)}, \vec Y_n,\m_I),
\ea
where $\Zv$, $\Zbf$, $\Z_{\text {fund}}$
and $\Z_{\text {afd}}$
denote  vector multiplet, 
bifundamental hypermultiplet, 
fundamental hypermultiplet  
and anti-fundamental hypermultiplet, respectively,
whose  explicit expressions are given in the appendix.
 $q=e^{\pi i \tau}$ is the instanton  expansion parameter.
$\vec a$ has  $N$ complex components
and represents the diagonalized vacuum expectation value
of vector multiplets.
$\m_I$ ($\tilde \m_I$,  $\n_{i}$) 
represents the mass of anti-fundamental
(fundamental,  bi-fundamental hypermultiplet). 
$\vec Y$ denotes the  $N$-tuple Young diagram 
$\vec Y=(Y_1,\cdots, Y_N)$. 

It is observed in \cite{BMZ_2015} that 
the instanton partition function can be  rewritten
as an  expectation value
\be
Z_{\text {inst}}^{(n+3)-point} 
=\langle G, \vec a, \tilde \m_I|\bigg\{\prod_{k=1}^{n-1}(q_{k}^D\,V_{k,k+1})\bigg\}
q_n^D| G, \vec a^{(n)}, \m_I; \vec M(n)\rangle
\label{expectation}
\ee
where  $D$ is an operator which counts 
 the number  of boxes in Young diagrams $|\vec{Y}|$,
and$V_{k,k+1}$ is the interwiner 
\begin{equation}\label{intertwiner}
V_{k,k+1}(\vec a^{(k)}, \vec a^{(k+1)}|\n_{k})
=\sum_{\vec Y_k,\vec Y_{k+1}}
\bZbf(\vec a^{(k)},\vec Y_k;\vec a^{(k+1)},\vec Y_{k+1}|\n_{k}) |\vec{a}^{(k)}, \vec{Y}_k
\rangle\langle \vec a^{(k+1)}+\n_{k}\vec e,\vec{Y}_{k+1}|
\end{equation}
where $\bZbf$ is rescaled using $\Zv$
\begin{equation}\label{intertwiner1}
\bZbf(\vec a^{(k)},\vec Y_k;\vec a^{(k+1)},\vec Y_{k+1}|\n_{k})
=\sqrt{\Zv(\vec a^{(k)},\vec Y_k)}~
\Zbf(\vec a^{(k)},\vec Y_k;\vec a^{(k+1)},\vec Y_{k+1}|\n_{k})
\sqrt{\Zv(\vec a^{(k+1)},\vec Y_{k+1})}.
\end{equation} 
The bra and ket in \eqref{intertwiner}
are the  AFLT bases  which satisfy the orthogonality and 
completeness  \cite{Alba2011}
\be
\langle \vec{a}, \vec{Y'}|\vec{a}, \vec{Y}\rangle=\d_{\vec Y',\vec Y},\quad 1=\sum_{\vec{Y}} |\vec{a},\vec{Y}\rangle \langle\vec{a},\vec{Y}|.
\label{AFLT-ortho}
\ee
In addition, the brackets  in \eqref{expectation} are  defined
on the AFLT basis 
\begin{align}
| G, \vec a^{(n)} , \m_I;
\vec M(n)\rangle
&=\sum_{\vec Y}\sqrt{\Zv(\vec a,\vec Y)}\prod_{I=1}^N 
\Z_{\text {afd}}(\vec a, \vec Y, \m_I) \,
|\vec a+ \vec M(n),\vec Y\rangle,
\label{G_ket}
\\ 
\langle G, \vec a, \tilde \m_I|&=\sum_{\vec Y}\sqrt{\Zv (\vec a,\vec Y)}\prod_{I=1}^N
 \Z_{\text {fund}}(\vec a, \vec Y, \tilde \m_I) \,\langle \vec a, \vec Y|\,.
\end{align}
Here, $\vec M(n)=\vec e\sum_{i=1}^{n-1}\n_i $, with  $\vec e=(1,1,...,1)$.  

One may evaluate the action of SH$^c$ generators on the basis $\aY$ 
based on the defining relations:
\begin{align}
\label{action}
D_{\pm1,n}\aY 
& =\sum_{x\in A/R(\vec{Y})}
(\phi_x)^n\L_x(\vec{Y})|\vec{a},\vec{Y}\pm x\rangle,
\quad D_{0,n+1}\aY=\sum_{x\in \vec{Y}}(\phi_x)^n\aY
\\
\langle \vec a, \vec Y|D_{\pm1,n}
&=\sum_{x\in R/A(\vec{Y})}
(\phi_x)^n\L_x(\vec{Y})\langle\vec{a},\vec{Y}\mp x|,\quad \langle \vec a, \vec Y|D_{0,n+1}=\sum_{x\in \vec{Y}}(\phi_x)^n\langle \vec a, \vec Y|
\end{align}
where the sets $A(\vec{Y})$ ($R(\vec{Y})$) 
contain all the boxes that can be added to 
(removed from) the Young diagram $\vec{Y}$ (see Figure 1). 
It is noted that the generator  of degree $\pm1$ 
adds/removes a box from the $N$-tuple Young diagram, 
which is denoted as $\vec{Y}\pm x$
following the convention used in  \cite{Bourgine2014a, BMZ_2015}.
The added/removed box $x$ is characterized by a triplet of indices
$(\ell;,i,j)$ where $\ell=1\cdots N$ and $(i,j)\in Y_\ell$ 
gives the position of the box in the $\ell$th Young diagram. 
To each box $x$ is associated with a complex number 
\be
 \phi_x=a_\ell+(i-1)\e_1+(j-1)\e_2 
\ee
and 
\be
\left\{  \L_x(\vec{Y})\right\}^{2}=\prod_{\superp{y\in A(\vec{Y})}{y\neq x}}\dfrac{\phi_x-\phi_y+\e_+}{\phi_x-\phi_y}\prod_{\superp{y\in R(\vec{Y})}{y\neq x}}\dfrac{\phi_x-\phi_y-\e_+}{\phi_x-\phi_y}\,.
\ee

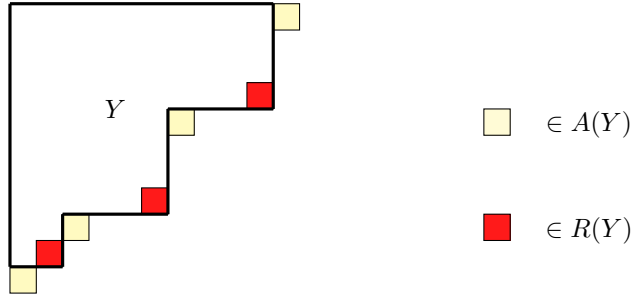
\begin{figure}
\begin{center}
\begin{tikzpicture}[scale=0.7]
\draw (2,-2) node {$Y$};
\filldraw [fill=yellow!30] (5,0) rectangle (5.5,-0.5);
\filldraw [fill=yellow!30] (3,-2) rectangle (3.5,-2.5);
\filldraw [fill=yellow!30] (1,-4) rectangle (1.5,-4.5);
\filldraw [fill=yellow!30] (0,-5) rectangle (0.5,-5.5);
\filldraw [fill=red!90] (4.5,-1.5) rectangle (5,-2);
\filldraw [fill=red!90] (2.5,-3.5) rectangle (3,-4);
\filldraw [fill=red!90] (0.5,-4.5) rectangle (1,-5);
\draw [very thick] (0,0) -- (5,0);
\draw [very thick] (5,0) -- (5,-2);
\draw [very thick] (5,-2) -- (3,-2);
\draw [very thick] (3,-2) -- (3,-4);
\draw [very thick] (3,-4) -- (1,-4);
\draw [very thick] (1,-4) -- (1,-5);
\draw [very thick] (1,-5) -- (0,-5);
\draw [very thick] (0,0) -- (0,-5);

\filldraw [fill=yellow!20] (9,-2) rectangle (9.5,-2.5);
\draw (11,-2.3) node {$ \in A(Y)$};
\filldraw [fill=red!90] (9,-4) rectangle (9.5,-4.5);
\draw (11,-4.3) node {$ \in R(Y)$};
\end{tikzpicture}
\end{center}
\caption{$A(Y)$ and $R(Y)$}
\end{figure}

The consistent condition of the action of the generators on AFLT basis 
results in the central charge of the form
$\cc_l=\sum_{p=1}^{N}(a_p +\e_+)^l$ \cite{Kanno:2013aha}. 
This identification shows that the central charge $c$
in \eqref{Hecke-CFT}  is given as  
\be
c=N - N(N^2-1){\e_+^2}/{ (-\e_1 \e_2)}.
\label{central-charge}
\ee
In addition, $J_0$ in \eqref{def:J_0} has an effect on $\aY$ as,
\be
J_0 \aY 
=\frac1{(-\e_1\e_2)}\bigg\{-\sum_{p=1}^{N}(a_p +\e_+) -N(N-1)\e_+ /2\bigg\}
\aY.
\ee

The Virasoro generator $L_0$ given  in \eqref{def:L_0} 
is defined in terms of $D_{0,1}$.  
According to \eqref{action} one may consider 
$D_{0,1}$ as the operator 
$D$ counting the number  $|\vec{Y}|$ of boxes:
 $D \aY= |\vec{Y}|\aY$. 
In this case, $L_0$ given  in \eqref{def:L_0} will have the form 
$ L_0=D+\Omega_0 $
where 
\be
\Omega_0 
=\bigg\{\sum_{p=1}^{N}(a_p +\e_+)^2
-\sum_{p=1}^{N}(a_p +\e_+)(1-N)\e_+ 
+N(N-1)(N-2)\e_+^2 /6  \bigg\}/(-2\e_1\e_2)\,.
\label{Omega0}
\ee
This shows that  $|\a, \vec 0 \rangle$ represents the primary state 
with conformal dimension $\Omega_0$
and $\aY$  a linear combination of Heisenberg+Virasoro descendants of total level $|\vec
 Y|$.

In order to compare later with the Liouville state, we define a modified  AFLT basis,
 \begin{equation}
 |\vec a,\vec Y,\delta_0\rangle= q_0^{\delta_0}|\vec a,\vec Y\rangle=|\vec a,\vec Y\rangle\otimes|\delta_0\rangle,
 \end{equation}
with $q_0 \to 0$, and introduce an operator $D_0=q_0\frac{\partial}{\partial q_0 } $, so that 
\begin{align}
D_0 |\vec a,\vec Y ; \d_0\rangle & =   \d_0   |\vec a,\vec Y ; \d_0\rangle,
\nn\\
D  |\vec a,\vec Y ; \d_0\rangle & =  |\vec{Y}|   |\vec a,\vec Y ; \d_0\rangle.
\end{align} 
Then, we shift $L_0 \to L_0 +D_0$ so that
\be
L_0  | \vec a, \vec Y; \d_0 \rangle  =( |\vec Y| + \d_0 + \Omega_0 ) 
| \vec a, \vec Y; \d_0 \rangle.
\label{new-L0}
\ee
$| \vec a, \vec Y; \d_0 \rangle $ is the primary state 
with conformal dimension $ (\d_0 + \Omega_0 ) $
and will  have an important role in investigating the AGT conjecture. We do not need shift $L_n$ for $n > 0$, since the corresponding  $q_0^{n+1}\frac{\partial}{\partial q_0 } $ term vanishes after taking $q_0 \to 0$.


Using the $q$-basis $| \vec a, \vec Y; \d_0 \rangle$, 
one may redefine the states shown  in \eqref{expectation}.
For example,
one may define $| G, \vec a, \m_I ; \d_0\rangle $ as in \eqref{G_ket}, 
where  $| \vec a, \vec Y \rangle $ 
is replaced with the modified AFLT basis  $| \vec a, \vec Y;\d_0 \rangle $
\be
\label{Gd}
| G, \vec a, \m_I ; \d_0\rangle
=\sum_{\vec Y}\sqrt{\Zv(\vec a,\vec Y)}\prod_{I=1}^N \Z_{\text {afd}}(\vec a, \vec Y, \m_I) \,|\vec a,\vec Y ; \d_0\rangle.
\ee
and
\be
\label{Gd-}
\langle G, \vec a, \tilde \m_I; \d_0|=\sum_{\vec Y}\sqrt{\Zv (\vec a,\vec Y)}\prod_{I=1}^N
 \Z_{\text {fund}}(\vec a, \vec Y, \tilde \m_I) \,\langle \vec a, \vec Y|q_0^{-\delta_0}\,.
\ee
And one may convert this state using $q$-basis by inserting $q_1^D$;
\be
|T_1  ; \d_0 \rangle
\equiv q_1^D| G, \vec a, \m_I  ; \d_0 \rangle
=  \sum_{\vec Y}  q_1^{|\vec Y|}  \sqrt{\Zv (\vec a,\vec Y)}\prod_{I=1}^N \Z_{\text {afd}}(\vec a, \vec Y, \m_I)   \,|\vec a,\vec Y; \d_0  \rangle.
\ee
Then the 4-punctured instanton partition function  \eqref{expectation}  
is written in terms of the new $q$-state; 
\be
\Z_{\text {inst}}^{4-point}
=\langle G, \vec a, \tilde \m_I ;\d_0| T_1   ; \d_0\rangle
=\langle G, \vec a, \tilde \m_I|q_1^D| G, \vec a, \m_I\rangle.
\ee
As far as the partition function is concerned, the $\d_0$ dependence 
canceled away. This means the partition function 
has the freedom to define the $q$-phase on the AFLT basis. Actually these 4 points
corresponds to the positions $0$, $q_1$, $1$ and $\infty$. Or equally, $q_0$, $q_1$, $1$ and $q_0^{-1}$, 
with $q_0 \to 0$. That's why we have explicit $q_0$ and $q_0^{-1}$ (and a hidden $1$) terms in 
\eqref{Gd} and \eqref{Gd-}.

Likewise, we define a $q$-state $|T_2  ; \d_0\rangle $ including one interwiner;
\begin{align}
|T_2  ; \d_0 \rangle 
&\equiv q_1^D ~V_{12}(\vec a, \vec b|\n)~q_2^D| G, \vec b, \m_I   ;\vec M ;\d_0(2)\rangle
\\
&= \sum_{\vec Y, \vec W}  q_1^{|\vec Y|}q_2^{|\vec W|}
\sqrt{\Z_{\text {vect}}(\vec a,\vec Y)}
\Zbf(\vec a,\vec Y;\vec b,\vec W|\n)
\Z_{\text {vect}}(\vec b,\vec W)
\prod_{I=1}^N \Z_{\text {afd}}(\vec b, \vec W, \m_I)   \,
|\vec a,\vec Y; \d_0(2) \rangle
\nn
\end{align}
where $\d_0(2)$ is put instead of $\d_0$ to emphasize that 
a different $q$-phase is used.  
Including $m-1$ interwiners, one has $|T_m ; \d_0\rangle$  
\be
|T_m  ; \d_0 \rangle
=   \bigg\{\prod_{k=1}^{m-1}\bigg(q_{k}^D\,
V_{k,k+1}(\vec a^{(k)}, \vec a^{(k+1)}|v_{k}) 
\bigg)\bigg\}q_m^D| G, \vec a^{(m)} , \m_I;\vec M(m); \d_0(m) \rangle\,. 
\ee
The instanton partition function is simply given as 
$ Z_{\text {inst}}^{(m+3)-point}=\langle G, \vec a, \tilde \m_I  ; \d_0| T_m  ; \d_0\rangle$. 
We will choose the $q$-phase $\d_0(m)$ as following (see  section \ref{sec:3.3} for details):
\be\label{delta0}
 \d_0 (m)={\frac{1}{{\e_1\e_2}}\bigg(\m_1\m_2+(\e_+-\sum_{i=1}^{m-1}\n_{i})(\m_1+\m_2)+\frac52\e_+^2+\sum_{r=1}^{m-1}(\e_+-\n_{r})
(3\e_+-\n_{r}-2\sum_{i=1}^{r-1}\n_{i})}\bigg)\,.
\ee 
 From now on, we will skip the notation $\d_0$ for simplicity,
assuming the AFLT basis is the $q$-basis and 
call $| T_m  \rangle$ gauge conformal state of rank $m$,
the counter part of the Liouville conformal state of rank $m$，
which will be considered in section \ref{sec:3.3}.

\section{ Construction of regular and irregular conformal states}

\subsection{Action of SH generators on $|T_m\rangle$ }
We summarize the results of actions of SH generators
on $|T_m\rangle$ using the $q$-differential representation,
the detailed calculation of which is shown in the appendix.
The non-trivial but a rather simple representation for $D_{0,1}$ 
is obtained if one identifies  $D_{0,1}=D+D_0$.
For any  state $|T_m\rangle$ of rank $m$, one has 
\be
D_{0,1} |T_m\rangle
=(D +D_0) |T_m\rangle
= \bigg[q_1\frac{\partial}{\partial q_1}+ \d_0(m)\bigg]|T_m\rangle\,.
\ee

A few operators
of order 0 and 1 are also shown, 
 which appear in  the defining relations in \eqref{SH1}-\eqref{SH4}.  
For rank 1 case, one has the representation 
 (same as the one  in  \cite{MRZ_2014}) 
\begin{align}
\label{rank101}
D_{-1,0}|T_1\rangle
&=q_1\frac{1}{\sqrt{-\e_1\e_2}}\sum_p^N (a_p +\m_p) |T_1\rangle\,,
\\
\label{rank111}
D_{-1,1}|T_1\rangle
&=
\bigg\{\sqrt{-\e_1\e_2}\,
q_1^2\frac{\partial}{\partial q_1}
+\frac12\frac{q_1}{\sqrt{-\e_1\e_2}}
\bigg [\sum_p^N\bigg( (a_p)^2-(\m_p)^2\bigg)
+\bigg(\sum_p^N ( a_p +\m_p)\bigg)^2~\bigg ] \bigg\}|T_1\rangle\,.
\\
\label{rank201}
D_{-2,0}|T_1\rangle
&=q_1^2 \sum_p (a_p +\m_p )|T_1\rangle,
\\
\label{rank211}
D_{-2,1}|T_1\rangle
&=
\left\{-2\e_1\e_2\,q_1^3\frac{\partial}{\partial q_1}+q_1^2
\Bigg [\sum_p^N\bigg( (a_p)^2-(\m_p)^2\bigg)+2\Bigg(\sum_p^N 
(a_p +\m_p)\bigg)^2\;\bigg ] \right\}|T_1\rangle\,
\end{align}

For rank 2, we have 
\begin{align}
D_{-1,0}|T_2\rangle
&= \bigg\{ q_1\frac{1}{\sqrt{-\e_1\e_2}}\sum_p^N(a_p-b_p+\e_+ -\n) +q_1q_2\frac{1}{\sqrt{-\e_1\e_2}}
\sum_p^N (b_p +\m_p)\bigg\} |T_2\rangle
\\
\label{rank112}
D_{-1,1}|T_2\rangle
&=
\bigg\{\sqrt{-\e_1\e_2}\,q_1(q_1\frac{\partial}{\partial q_1}-q_2\frac{\partial}{\partial q_2})+
{\frac12}\frac{q_1}{\sqrt{-\e_1\e_2}}\bigg [\sum_p^N\bigg(a_p^2-(b_p-\e_+ +\n)^2\bigg)
+\bigg(\sum_p^N(a_p-b_p+\e_+ -\n)\bigg)^2~\bigg ]
 \nn\\
&+\sqrt{-\e_1\e_2}\,
q_1q_2(q_2\frac{\partial}{\partial q_2})
+\frac12 \frac{q_1q_2}{\sqrt{-\e_1\e_2}}
\bigg [\sum_p^N\bigg( (b_p+\n)^2-(\m_p-\n)^2\bigg)
+ \bigg(\sum_p^N(  b_p+\m_p)\bigg)^2~\bigg ]\bigg\}
|T_2\rangle\,,
\\
D_{-2,0}|T_2\rangle
&=\bigg\{q_1^2\sum_p^N (a_p-b_p+\e_+ -\n)+
q_1^2q_2^2\sum_p^N \big(b_p +\m_p\big)\bigg\}|T_2\rangle\,,
\\
\label{rank212}
D_{-2,1}|T_2\rangle
&=
\bigg\{-2\e_1\e_2q_1^2
(q_1\frac{\partial}{\partial q_1}-q_2\frac{\partial}{\partial q_2})
+q_1^2\bigg [\sum_p^N\bigg(a_p^2-(b_p-\e_+ +\n)^2\bigg)
+2\bigg(\sum_p^N (a_p-b_p+\e_+ -\n)\bigg)^2\bigg ]\\
&-2\e_1\e_2q_1^2q_2^2(q_2\frac{\partial}{\partial q_2})
+q_1^2q_2^2\bigg [\sum_p^N\bigg( (b_p+\n)^2-(\m_p-\n)^2\bigg)
+2\bigg(\sum_p^N b_p+\m_p )\bigg)^2~ \bigg ]\bigg\}|T_2\rangle\,. \nn
\end{align}
For rank $m$,  we find
\begin{align}
D_{-1,1}|T_m\rangle
&=
\sum_{k=1}^{m-1}\frac{q_1\cdots q_k}{\sqrt{-\e_1\e_2}}\bigg\{-\e_1\e_2(q_k\frac{\partial}{\partial q_k}-q_{k+1}\frac{\partial}{\partial q_{k+1}})
\\
+&{\frac12}\sum_p^N\bigg[(a_{p}^{(k)} +\sum_{i=1}^{k-1}\n_{i})^2-(a_{p}^{(k+1)} -\e_+ +\sum_{i=1}^{k}\n_{k})^2\bigg] 
+{\frac12}\bigg(\sum_p^N (a_{p}^{(k)}-a^{(k+1)}_{p}+\e_+ -\n_{k})\bigg)^2 \bigg\}
\nn\\
+&\frac{q_1\cdots q_m}{\sqrt{-\e_1\e_2}}\bigg\{ -\e_1\e_2q_m\frac{\partial}{\partial q_m}+\frac12\sum_p^N \bigg((a^{(m)}_{p} +\sum_{i=1}^{k-1}\n_{i})^2-(\m_p-\sum_{i=1}^{k-1}\n_{i})^2\bigg)+\frac12(\sum_p a^{(m)}_{p}+\m_p)^2 \bigg\}|T_m\rangle\nn\,.
\\
D_{-2,1}|T_m\rangle
&=
\sum_{k=1}^{m-1}(q_1\cdots q_k)^2\bigg\{-2\e_1\e_2(q_k\frac{\partial}{\partial q_k}-q_{k+1}\frac{\partial}{\partial q_{k+1}})
\\
+&\sum_p^N\bigg[(a_{p}^{(k)} +\sum_{i=1}^{k-1}\m_{i,i+1})^2-(a_{p}^{(k+1)} -\e_+ +\sum_{i=1}^{k}\n_{k})^2\bigg]
+\bigg(\sum_p^N (a_{p}^{(k)}-a^{(k+1)}_{p}+\e_+ -\n_{k})\bigg)^2\bigg\}
\nn\\
+&(q_1\cdots q_m)^2\bigg\{ -2\e_1\e_2q_m\frac{\partial}{\partial q_m}+\sum_p^N \bigg((a^{(m)}_{p} +\sum_{i=1}^{k-1}\n_{i})^2-(\m_p-\sum_{i=1}^{k-1}\n_{i})^2\bigg)+2(\sum_p a^{(m)}_{p}+\m_p)^2 \bigg\}|T_m\rangle\nn.
\end{align}
\subsection{Virasoro  action on  $|T_m\rangle$}

In this section, we provide the $q$-representation of the 
Virasoro generators. For this purpose, we restrict
ourselves to the gauge group  SU$(2)$ 
for which we put $N=2$ and further  require $\sum_p(a_{p}^{(k)})=0$.

The $L_0$ defined in \eqref{new-L0} 
has the $q$-differential representation on the gauge 
conformal state of rank $m$ 
\be\label{L0Tm}
L_0|T_m\rangle=\left( q_1\frac{\partial}{\partial q_1}+ \d_0+\Omega_0
\right) |T_m\rangle\,,
\ee
where $\Omega_0$ defined in \eqref{Omega0}
has a simple form $\Omega_0= \bigg( \frac12\sum_pa_p^2+2\e_+^2 \bigg)/({-\e_1\e_2})$.

According to \eqref{defV}, we have 
$L_1=-\frac{1}{\sqrt{-\e_1\e_2}}D_{-1,1}$ 
and $L_2=-\frac{1}{2\e_1\e_2}D_{-2,1}-\frac{\e_+}{\e_1\e_2}D_{-2,0}$
in terms SH generators. 
Therefore, using the results in section 3.1, we have on the state of  rank 1
\begin{align}
\label{L1T1}
L_1|T_1\rangle
&=q_1^2\frac{\partial}{\partial q_1}
+\frac{q_1}{{-\e_1\e_2}}
\bigg\{ \frac12\sum_p (a_p)^2+\m_1\m_2 \bigg\},
\\
\label{L2T1}
L_2|T_1\rangle
&=q_1^3\frac{\partial}{\partial q_1}+\frac{q_1^2}{{-\e_1\e_2}}\bigg\{  \frac12\sum_p (a_p)^2+\m_1\m_2+ \frac12(\m_1+\m_2) ^2+{\e_+}(\m_1+\m_2) \bigg\}.
\end{align}
On $|T_2\rangle$ we have 
\begin{align}
\label{L1T2}
L_1|T_2\rangle
&=
\bigg\{\,q_1(q_1\frac{\partial}{\partial q_1}-q_2\frac{\partial}{\partial q_2})+\frac{q_1}{-\e_1\e_2}\bigg [{\frac12}\sum_p\bigg(a_p^2-b_p^2\bigg)+ (\e_+ -\n)^2\bigg ]\\
&~~~+q_1q_2(q_2\frac{\partial}{\partial q_2})+\frac{q_1q_2}{-\e_1\e_2}\bigg [\frac12\sum_p b_p^2+\m_1\m_2+\n(\m_1+\m_2)\bigg ]\bigg\}|T_2\rangle\,, \nn
\\
\label{L2T2}
L_2|T_2\rangle
&=
\bigg\{q_1^2(q_1\frac{\partial}{\partial q_1}-q_2\frac{\partial}{\partial q_2})+\frac{q_1^2}{-\e_1\e_2}\bigg [{\frac12}\sum_p\bigg(a_p^2-b_p^2\bigg)+ 3(\e_+ -\n)^2+2\e_+(\e_+ -\n)\bigg ]\\
&+q_1^2q_2^2(q_2\frac{\partial}{\partial q_2})+\frac{q_1^2q_2^2}{-\e_1\e_2}\bigg [\frac12\sum_p b_p^2+\m_1\m_2+\n(\m_1+\m_2)+ \frac12(\m_1+\m_2)^2+\e_+(\m_1+\m_2)\bigg ]\bigg\}|T_2\rangle\,. \nn
\end{align}
The same method applies to $|T_m\rangle$ 
\begin{align}
\label{L1Tm}
L_1|T_m\rangle
&=
\sum_{k=1}^{m-1}\frac{q_1\cdots q_k}{-\e_1\e_2}\bigg\{-\e_1\e_2(q_k\frac{\partial}{\partial q_k}-q_{k+1}\frac{\partial}{\partial q_{k+1}})+(\e_+-\n_{k}+2\sum_{i=1}^{k-1}\n_{i})(\e_+-\n_{k})
\\
&+{\frac12}\sum_p (a^{(k)}_{p})^2-{\frac12}\sum_p (a^{(k+1)}_{p})^2\bigg\}
+\frac{q_1\cdots q_m}{-\e_1\e_2}\bigg\{ -\e_1\e_2q_m\frac{\partial}{\partial q_m}+\frac12\sum_p (a^{(m)}_{p} )^2+\m_1\m_2+(\m_1+\m_2)\sum_{i=1}^{m-1}\n_{i} \bigg\}|T_m\rangle\nn
\\
\label{L2Tm}
L_2|T_m\rangle
&=\sum_{r=1}^{m}(q_1\cdots q_r)^{2}
\frac{\partial}{\partial (q_1\cdots q_r)}\\
&+
\sum_{k=1}^{m-1}\frac{(q_1\cdots q_k)^2}{-\e_1\e_2}\bigg\{{\frac12}\sum_p (a^{(k)}_{p})^2-{\frac12}\sum_p (a^{(k+1)}_{p})^2+(5\e_+-3\n_{k}+2\sum_{i=1}^{k-1}\n_{i})(\e_+-\n_{k})\bigg\}\nn\\
&+\frac{(q_1\cdots q_m)^2}{-\e_1\e_2}\bigg\{ \frac12\sum_p (a^{(m)}_{p} )^2+\m_1\m_2+(\m_1+\m_2)(\e_++\sum_{i=1}^{m-1}\n_{i})+\frac12\m_1^2+\frac12\m_2^2 \bigg\}\nn
\end{align}

\subsection{Comparison of gauge conformal state $|T_m\rangle$ with the Liouville state  $\vert R_m \rangle$}
\label{sec:3.3}

One may define a conformal state $\vert R_m \rangle$ by
applying an $m$-product of  primary  fields $\Psi_{\Delta_r}(z_r) $ 
of conformal dimension $\Delta_r$ at positions $z_r$ 
on a  primary state 
$ \vert \Delta_{0} \rangle$  of conformal dimension $ \Delta_{0} $;
\be\label{rmrm}
\vert R_m \rangle= \prod_{r=1}^m \Psi_{\Delta_r}(z_r) \vert \Delta_{0} \rangle.
\ee 
We will use the (imaginary) Liouville vertex operator as a primary field
with  $N$ free fields $ \vec\varphi=(\varphi_1, \cdots, \varphi_N)$;
$ \Psi_{\Delta_r}(z_r) 
= e^{i\vec \k^{(r)} \cdot \vec\varphi(z_r)}$ 
which has the conformal dimension 
\be
\D_r=\sum_i^N\frac12\k_i^{(r)}(\k_i^{(r)}+2Q\rho_i)
\label{Liouv-dim}
\ee
where  the component notation for $\vec \k^{(r)}=(\k_1, \cdots, \k_N)$ is used and  $Q\rho_i$ is the background charge.
The primary state is defined as 
$ \vert \Delta_{0} \rangle=\lim_{z_0 \to 0} \Psi_{\Delta_0}(z_0) \vert 0 \rangle$. 
Then, the Virasoro generator $\cL_k$ with $  (k\geq -1) $  
 has the holomorphic representation  on the conformal state 
\begin{equation}\label{LRz}
\cL_k|\,R_m\,\rangle\,=\,\sum_{r=0}^{m}z_r^k
\left(z_r\frac{\partial}{\partial z_r}+(k+1) \D_r\right)|\,R_m\,\rangle\,.
\end{equation}

The holomorphic state $|\,R_m\,\rangle$
and the gauge conformal state $|T_m\rangle $
have similar structures and 
their parameters are identified with each other. 
If one compares  the Virasoro action on 
 $|\,R_1\,\rangle$ with $|T_1\rangle $
using the relations given in \eqref{L0Tm}, \eqref{L1T1}, 
 \eqref{L2T1} and \eqref{LRz},
one can equate $z_1$ with $ q_1$.
However, there is a slight mismatch between 
$|\,R_1\,\rangle$  and $|T_1\rangle $.
To fix this, one needs to modify $|\,T_1\,\rangle$ 
by multiplying a function of $q_1$
and  finds 
\be
\label{T1R1}
|\,K_1\,\rangle\,=q_1^{-F_1(1)}|T_1\rangle 
\ee
where $F_1(1)=\Big(\frac12\sum_p (a_p)^2+\m_1\m_2-(\m_1+\m_2)^2-2\e_+(\m_1+\m_2) \Big)/ (\e_1\e_2)$. 
Then, using \eqref{T1R1} and \eqref{L0Tm}  
one finds  $L_0$  on  $|\,K_1\,\rangle $,
\be\label{L0K1}
L_0
|\,K_1\,\rangle\,=\left\{q_1\frac{\partial}{\partial q_1}+ \d_0+\Omega_0 
+ F_1(1) \right\}  |K_1\rangle .
\ee
This is compatible with \eqref{LRz}  if one requires 
$\d_0$ to have the form
\be
\d_0=\D_1+\D_0 -(  F_1(1) + \Omega_0).
\ee
In addition, the actions of $L_1$ and $L_2$ on $|\,K_1\,\rangle$ 
provide the relations between other parameters
which is summarized as follows.
The background charge in \eqref{Liouv-dim}  is given as 
\be
\label{background_Q}
Q = {-\sqrt2 \e_+}/{\sqrt{-\e_1\e_2}}
\ee 
so that  $\rho_1=-\rho_2=-\frac12 $.  
This shows that the central charge in \eqref{central-charge}
is given as $c=1-3Q^2$ for SU(2) gauge group. 
Holomorphic coordinates are identified as 
$ z_1= q_1$ and $ z_{0}= 0$ and 
conformal dimensions are given as 
\begin{align} 
\k_1^{(1)}& =-\k_2^{(1)}= \frac{\m_1+\m_2}{\sqrt{-2\e_1\e_2}}, ~~
\k_1^{(0)}=-\k_2^{(0)}=\frac{Q}{2}+ \frac{\m_1-\m_2}{\sqrt{-2\e_1\e_2}},
\\
\D_1&=\frac{1}{{-\e_1\e_2}}\bigg\{  \frac12(\m_1+\m_2) ^2+{\e_+}(\m_1+\m_2) \bigg\},
~~
\D_0=\frac{1}{{-\e_1\e_2}}\bigg\{  \frac12(\m_1-\m_2) ^2- \frac12\e_+^2 \bigg\}  \,.
\end{align}
This parameter identification leads to $\d_0=\d_0(1)$ as given in \eqref{delta0}
where we use the relation $ F_1(1) + \Omega_0 = 
\Big( -\m_1\m_2+(\m_1+\m_2)^2+2\e_+(\m_1+\m_2)+2\e_+^2 \Big)/ {(-\e_1\e_2) }$.

Note that $L_1$ and $L_2$ are enough to generate the full (positive)
Virasoro algebra by commutation relations
and all of $L_k$'s action on $|T_1\rangle$ or $|\,K_1\,\rangle$ are fixed. 
This  demonstrates that the state $|K_1\rangle$
constructed from  the gauge theory side 
is equivalent to  the state $|\,R_1\,\rangle$ constructed 
from the Liouville vertex operators. 

This identification procedure can be generalized to higher rank case. 
In the same way as rank 2, we find that 
\eqref{L0Tm}, \eqref{L1T2} and \eqref{L2T2}
are consist with their Liouville conformal counterparts \eqref{LRz}, 
as long as $|K_2\rangle$ is identified with $|\,T_2\,\rangle$ with a prefactor, 
\begin{align}
\label{T2R2}
|K_2\rangle &=q_1^{-F_1(2)}q_2^{-F_2(2)}  |\,T_2\,\rangle,
\\
F_1(2)&=  \frac{1}{{\e_1\e_2}}
\left(\frac12\sum_p (a_p)^2-\frac12\sum_p (b_p)^2-3(\e_+- \n)^2-4\e_+(\e_+- \n) \right)+ F_2(2),
\nn\\
F_2(2)&= \frac{1}{{\e_1\e_2}}\left(\frac12\sum_p (b_p)^2+\m_1\m_2-(\m_1+\m_2)^2-2\e_+(\m_1+\m_2)+\n(\m_1+\m_2) \right) .
\nn
\end{align} 
The prefactor allows the differential representation 
\be
L_0 |\,K_2\,\rangle\,=
\left( q_1\frac{\partial}{\partial q_1}+ \d_0+ \Omega_0+ F_1 (2)\right)|K_2\rangle.
\ee
Noting the relations  $z_1= q_1$, $z_2= q_1 q_2$ and $z_0= 0$,   
we have 
\be
 \d_0+ \Omega_0+ F_1 (2) = \D_0+ \D_1 + \D_2.
\label{F1(2)}
\ee 
If one incorporates $L_1$ and $L_2$, 
one has $\d_0=\d_0(2)$ as  in \eqref{delta0}.
The background charge $Q$ is the same as that in \eqref{background_Q}
and conformal dimensions  are given as  
\begin{align}
\k_1^{(2)}&=-\k_2^{(2)}= \frac{\m_1+\m_2}{\sqrt{-2\e_1\e_2}}\,,\qquad
\k_1^{(1)}=-\k_2^{(1)}=-Q- \frac{\sqrt2 \n}{\sqrt{-\e_1\e_2}}\,,\qquad
\k_1^{(0)}=-\k_2^{(0)}=\frac{Q}{2}+ \frac{\m_1-\m_2}{\sqrt{-2\e_1\e_2}}\,, 
\\
\D_2 &=\frac{1}{{-\e_1\e_2}}\bigg\{  \frac12(\m_1+\m_2) ^2+{\e_+}(\m_1+\m_2) \bigg\}\,, \nn\\
\D_0&=\frac{1}{{-\e_1\e_2}}\bigg\{  \frac12(\m_1-\m_2) ^2-\frac12\e_+^2 \bigg\}\,, \qquad
\D_1=\frac{1}{{-\e_1\e_2}}\bigg\{  2(\e_+ -\n) ^2+2{\e_+}(\e_+ -\n) \bigg\}  \,.
\end{align}

It is straight-forward to compare \eqref{L0Tm}, \eqref{L1Tm} and \eqref{L2Tm} with \eqref{LRz} once the prefactor is found.  
\ba\label{TmRm}
|K_m\rangle&&=\prod_{r=1}^{m-1}(q_1\cdots q_r)^{\frac{1}{{\e_1\e_2}}\big({\frac12}\sum_p^2 (a^{(r)}_{p})^2-{\frac12}\sum_p^2 (a^{(r+1)}_{p})^2+(3\n_{r}-7\e_++2\sum_{i=1}^{r-1}\n_{i})(\e_+-\n_{r}) \big)}\nn\\
&&\times(q_1\cdots q_m)^{\frac{1}{{\e_1\e_2}}\big(\frac12\sum_p (a^{(m)}_{p})^2+(\m_1+\m_2)(-2\e_++\sum_{i=1}^{m-1}\n_{i})+\m_1\m_2-(\m_1+\m_2)^2\big)}
|\,T_m\,\rangle.
\ea
The holomporphic coodinates are identified with the $q$-coordinates
\be
z_r = q_1\cdots q_r\,,\qquad(r=1, \cdots m);\qquad
z_{0}= 0.
\ee  
The background charge $Q$ is the one in \eqref{background_Q}
and $\d_0=\d_0(m)$ in \eqref{delta0}.
Conformal dimensions are given as  
\begin{align} 
\k_1^{(r)} &=-\k_2^{(r)}=-Q- \frac{\sqrt2 \n_{r}}{\sqrt{-\e_1\e_2}}\,,\quad
(1\leq r \leq m-1)
\nn\\
\k_1^{(m)} &=-\k_2^{(m)}= \frac{\m_1+\m_2}{\sqrt{-2\e_1\e_2}}\,,\quad
\k_1^{(0)}=-\k_2^{(0)}=\frac{Q}{2}+ \frac{\m_1-\m_2}{\sqrt{-2\e_1\e_2}},
 \\
\D_r &=\frac{1}{{-\e_1\e_2}}\bigg\{  2(\e_+ -\n_{r}) ^2+2{\e_+}(\e_+ -\n_{r}) \bigg\},\quad
(1\leq r \leq m-1) ,
\nn\\
\D_m &=\frac{1}{{-\e_1\e_2}}\bigg\{  \frac12(\m_1+\m_2) ^2+{\e_+}(\m_1+\m_2) \bigg\}\,,\qquad
\D_0=\frac{1}{{-\e_1\e_2}}\bigg\{  \frac12(\m_1-\m_2) ^2-\frac12\e_+^2 \bigg\} .
\end{align}
These parameter relations are  exactly the AGT dictionary, which
translates the CFT parameters to their gauge counterparts.
The (imaginary) Liouville CFT side is based on one boson construction, 
with the vertex operator $ \Psi_{\tilde \Delta_r}(z_r) = e^{i \a^{(r)}  \psi(z_r)}$ and
conformal dimension $\tilde \D_r\equiv \a^{(r)}(\a^{(r)}-Q)$
if one uses the relation with our two boson $\varphi^{(i)}$ construction
$\psi=\frac12\varphi_1-\frac12\varphi_2$ and $\a_r=\k_1^{(r)} =-\k_2^{(r)}$.

\section{Colliding limit and irregular state}
 Virasoro representation ${\mathcal{L}}_k$ on  the irregular state
 $\vert I_m \rangle$ of rank $m$ 
is given in terms of differential operators 
with respect to the eigenvalue $c_k$ 
of positive mode of Heisenberg operator  $a_k$
with $0 \le k \le m$ \cite{GT2012} 
\be
\label{virasoro}
{\mathcal{L}}_k =
\begin{cases} 
 \L_k +\sum_{l=1}^{m-k}l\,c_{l+k}\frac{\partial}{\partial c_{l}}
&{\rm for}~0\le k \le 2m \\
0  &{\rm for}~ k >2m   
\end{cases}
\ee 
where 
$\Lambda_k =\sum_{l} c_{l}c_{k-l} -(k+1)Qc_k$.
It is noted that if  $\cL_1$  and $\cL_2$ are given,
then other generators in \eqref{virasoro} are determined 
from the Virasoro commutation relations.
In addition, ${\mathcal{L}}_k $  with $m\le k \le 2m$ 
reduces to the eigenvalue  $\L_k$ since there is no $c_k$ with $k>m$. 
Therefore, when the stress energy tensor applies on the irregular state of rank $m$, 
one has singular contributions
\begin{equation}
T_>(y)|\,I_m\,\rangle\,=\,\Biggl[\;
\sum_{k=m}^{2m}\frac{\Lambda_k}{y^{k+2}}+
\sum_{k=0}^{m-1}\frac{{\mathcal{L}}_k}{y^{k+2}}+\frac{1}{y}L_{-1}\,\Biggr]
|\,I_m\,\rangle\,.
\end{equation}
The irregular state is of the form \cite{KMST2013} 
\be
|  I_{m} \rangle =
\sum_{\ell,Y, \ell_p} 
\L^{\ell/m}
\left\{ 
\prod_{i=1}^{m-1}  a_i ^{\ell_{2m-i} } b_i^{\ell_i}  
\right\} 
t^{\ell_m} 
Q_{\Delta} ^{-1} \Big(
1^{\ell_1}2^{\ell_2}
\cdots 
 (2m-1) ^{\ell_{2m-1}} (2m) ^{\ell_{2m}}; Y \Big)
L_{-Y} |\Delta \rangle 
\label{G_2m}
\ee
where $\ell=|Y|$. The eigenvalues are $\Lambda_m=\Lambda t$,
$\Lambda_{2n-s}=\Lambda^{(2n-s) /n} a_s $ and $\Lambda_{2n}=\Lambda^{2}$.

To obtain the colliding limit from the regular state 
we need to scale away the singular contribution,
which is achieved if one defines $|\, R_1'\,\rangle$  as 
\begin{equation}
|\,R_1'\,\rangle\,=\,z_1^{-2\a_1\a_0}\,|\,R_1\,\rangle
\label{R1_R1'}
\end{equation} 
since Virasoro generators
has the differential representation on $|\,R_1'\,\rangle$  
\begin{align}
\label{rank1L}
\cL_0|R_1'(z)\rangle
&= 
\left( z_1\frac{\partial}{\partial z_1}+(\a_1+\a_0) (\a_1+\a_0-Q) \right)
|R_1'(z)\rangle
\\
\cL_1|R_1'(z)\rangle
&= 
\left( z_1^2\frac{\partial}{\partial z_1}+2z_1\a_1(\a_1+\a_0-Q) \right) 
|R_1'(z)\rangle
\nn\\
\cL_2|R_1'(z)\rangle\,
&=\left(
z_1^3\frac{\partial}{\partial z_1}+z_1^2\a_1(3\a_1+2\a_0-3Q)\right)
|R_1'(z)\rangle. 
\nn
\end{align}

On the other hand, according to \eqref{L0K1}, the gauge conformal state has the form
\be
L_0
|\,K_1\,\rangle\,=\bigg\{q_1\frac{\partial}{\partial q_1}+\D_1+\D_0\bigg\}|K_1\rangle .
\ee
However, considering the fusion of two vertex operators at $z_1$ 
and the origin, we need a $q$-state $|\,K_1'\,\rangle$ obeying
\be
L_0
|\,K_1'\,\rangle\,=\bigg\{q_1\frac{\partial}{\partial q_1}+\D_{01}\bigg\}|K_1'\rangle 
\ee
where $\D_{01}=\a_{01}(\a_{01}-Q)$  with $\a_{01}=\a_0+\a_1$
as given in \eqref{rank1L}.
This is achieved if a new $q$-representation
$|\,K_1'\,\rangle$ is defined as
\be 
\label{K1K1'}
|\,K_1'\,\rangle\,=
  q_1^{(\D_0+\D_1)-\D_{01}}|K_1\,\rangle=\,q_1^{-2\a_1\a_0}\,|K_1\,\rangle=\,q_1^{-H_1}\,|T_1\,\rangle,
\ee
where $ H_1=
 [ \D_{01}- (\D_1+\D_0)  ] + F_1
=[\D_{01}-(\D_1+\D_0)]  +[\D_1+\D_0- (\d_0(1)+\Omega_0)]$ 
and its explicit value is given as
$H_1= 
{\frac{1}{{-\e_1\e_2}}\big(\frac12\sum_p (a_p)^2-(\m_1+\e_+)(\m_1+\m_2)-\m_1^2 \big)} $. 

The colliding limit is to put 
$\a_i \to \infty$ and $z_i \to 0$ 
while keeping $c_1=z_1\a_1$ and $c_0=\a_1+\a_0$  finite
and reduces $|\,R_1'\,\rangle$ to $|\,I_1\,\rangle$.
\be
\cL_1|\,I_1 \rangle\,
=2c_1 (c_0-Q)|\,I_1\,\rangle\,, ~~ 
\cL_2|\,I_1 \,\rangle\,
= c_1^2 |\,I_1 \,\rangle.
\ee 
Since the actions of $\cL_1$ and $\cL_2$ commute each other, 
$\cL_k=0$ when $k \ge 3$. 

On the same footing,
$|K_1' \rangle$ becomes the irregular state of rank 1 
since $|K_1' \rangle$ and  $|R_1' \rangle$ have the 
same differential structure  when $z_1=q_1$. 
Therefore, we may obtain  $|I_1 \rangle$ 
in terms of $|K_1' \rangle$ at the colliding limit up to normalization,
if $|\D \rangle$ in \eqref{G_2m}
is identified with the newly defined $q$-basis 
$|\D \rangle  \equiv  
 \lim_{q_1 \to 0}(\a_1q_1)^{-H_1}|\,\vec a, 0; \d_0(1)\,\rangle$.
Its descendant  
$|\D+Y \rangle =L_{-Y}|\D \rangle $ 
is given as 
$ c_1^{-H_1}|\,\vec a, \vec Y; \d_0(1)\,\rangle$, since 
$L_0|\D+Y \rangle=(\D+|Y|)|\D+Y \rangle$.
After this consideration, 
the irregular conformal state of rank 1 in \eqref{G_2m}
is written in terms of the $q$-basis  as appeared in \cite{MRZ_2014} 
\be
|I_1\rangle=\sum_{\vec Y}\L^{|\vec Y|}(Z_{\text {vect}}(\vec a,\vec Y)
)^{1/2}Z_{\text {afd}}(\vec a, \vec Y, \m_1) |\D+Y \rangle
\ee 
which  is equivalent to  the one given in  \cite{MMM2009} 
when the colliding limit is achieved with  
$\m_2\to \infty, q \to 0$ and $q\m_2=\L$ finite. 
As a result, the inner product 
$\langle \D |I_1\rangle =\langle \D | \D \rangle$, 
which can be normalized as 1. 

The irregular state of rank 2 can be constructed similarly. 
First, we  prepare $|\,R_2'\,\rangle$  for the colliding limit:
\be 
|\,R_2'\,\rangle\,=\,z_1^{-2\a_1\a_0}z_2^{-2\a_2\a_0}(z_1-z_2)^{-2\a_1\a_2}\,|\,R_2\,\rangle\,.
\label{R2R2'}
\ee
The Virasoro representation is given as 
\begin{align}
\label{rank2L}
\cL_0|R_2'(z)\rangle
&= \left( z_1\frac{\partial}{\partial z_1 } +z_2\frac{\partial}{\partial z_2 } 
+\D_{012}  \right) |\,R_2'(z)\rangle
\\
\cL_1|R'_2(z)\rangle
&= \sum_{a=1,2} \left( z_a^2\frac{\partial}{\partial z_a}
 +2z_a\a_a(\a_{012} -Q) \right) |R'_2(z)\rangle
\nn\\
\cL_2|R'_2(z) \rangle\,
&=
\left(
 z_1^3\frac{\partial}{\partial z_1}  + z_2^3\frac{\partial}{\partial z_2} 
+z_1^2\a_1(\a_1+2\a_{012} -3Q)
+z_2^2\a_2(3Q-2\a_{012} -\a_2)+2z_1\a_1z_2\a_2\right)
|R'_2(z) \rangle\,,
\nn
\end{align}
where $\D_{012} = (\a_{012}) (\a_{012}-Q)$
and $\a_{012}=\a_2+ \a_1+\a_0$.
At the colliding limit we have finite variables 
$c_2=z_1^2\a_1+z_2^2\a_2$, $c_1=z_1\a_1+z_2\a_2$ and $c_0=\a_1+\a_2+\a_0$
and the Virasoro operation reduces to  
\begin{align}
\cL_0|I_2 \rangle
&= \left( c_1\frac{\partial}{\partial c_1 } +2 c_2\frac{\partial}{\partial c_2 } 
+\D_{012}  \right) |\,I_2 \rangle
\\
\cL_1|\,I_2\,\rangle\,
&= \left(c_2\frac{\partial}{\partial c_1}+2c_1(c_0-Q) \right) |I_2 \,\rangle\,,
\\
\cL_2|I_2 \,\rangle\,
&= \Big( (2c_0-3Q)c_2+c_1^2 \Big)|I_2 \rangle\,.
\end{align}
Higher positive non-vanishing generators 
$\cL_3$ and $\cL_4$
are generated from 
the lower generators $ \cL_1$ and $\cL_2$
and the irregular state of rank 2 is the simultaneous 
eigenstate of the three generators $\cL_2$, $\cL_3$ and $\cL_4$.  

Similarly for the $q$-state of rank 2,  we have  
\be
L_0
|K_2 \rangle\,=\bigg\{q_1\frac{\partial}{\partial q_1}+\D_2+\D_1+\D_0\bigg\}
|K_2\rangle .
\ee
However, we need a state $|K_2' \rangle$
\be
L_0
|K_2' \rangle 
=\bigg\{q_1\frac{\partial}{\partial q_1}+\D_{012}\bigg\}
|K_2 '\rangle  
\ee
which can be realized if one puts
\be
|\,K_2' \rangle =q_1^{(\D_0+\D_1+\D_2)-\D_{012}} h(q_2)|K_2 \rangle.
\label{K1K1-1}
\ee 
Considering the identification for the regular conformal state,
we may have the form 
\be
\label{K2K2'}
|K_2' \rangle
= (q_1)^{-2\a_1\a_0} (q_1q_2)^{-2\a_2\a_0}(q_1-q_1q_2)^{-2\a_1\a_2} 
|K_2\rangle 
\ee
if one uses the relation $z_1=q_1$ and $z_2= q_1 q_2$
and \eqref{R2R2'}. 
It is easy to convince that  $q_1$ power
in \eqref{K1K1-1} matches with the one in \eqref{K1K1'}:  
\be
\D_{012} -(\D_0+\D_1+\D_2) = 2\a_1\a_0 + 2\a_2\a_0 + 2\a_1\a_2. 
\ee
The remaining factor $h(q_2)$ in \eqref{K1K1-1} 
can be fixed as 
\be
h(q_2) = q_2^{-2\a_2\a_0}(1-q_2)^{-2\a_1\a_2}.
\label{h(q2)}
\ee

Note that the finite parameters at the colliding limit 
are related with  $q_1$ and $q_2$
\be
c_0=\a_0 + \a_1+\a_2,
~~~~
c_1=q_1\a_1(1 + q_2\a_2/\a_1), 
~~~~
c_2 = q_1^2 \a_1 (1 + q_2^2 \a_2/\a_1) .
\ee
Finite $c_2$ is obtained at the colliding limit 
as $\a_1 \to \infty$ and $q_1 \to 0$.
Therefore, we may ask  if 
$q_1^2 \a_1$  and $ q_2^2 \a_2/\a_1$ are separately 
finite. 
This is the case both $q_1$ and $q_2$ go to 0
because as $\a_1$ goes to infinity, 
so does  $(\a_0 + \a_2)$.
Therefore, $q_2 \to 0$ limit ensures that 
  $ q_2^2 \a_2/\a_1$ is finite
since the ratio $a_2/a_1$ can be infinite. 
On the other hand,  the limit $q_2 \to 1$
is not allowed in the colliding limit
because as  $q_2 \to 1$
we have $c_1 \to q_1\a_1(1 +\a_2/\a_1)$
 and $c_2  \to  q_1^2 \a_1 (1 +  \a_2/a_1)$ 
which cannot simultaneously be finite (non-zero) as $q_1 \to 0$.

In fact, the limit $q_2 \to 1$ corresponds to 
the  t-channel limit $z_1 \sim z_2 \to 0$
and is not allowed in the colliding limit.  
In contrast, the limiting procedure  $q_1, q_2 \to 0$ 
corresponds to the limit 
$|z_2| <|z_1 | \to 0$.
This concludes that the colliding limit 
allows only the s-channel limiting procedure
and the hierarchical behavior $z_2 <z_1 \to 0$
(or $q_1, q_2 \to 0$) 
should be present in the final result.  
More explicitly, this s-channel limit
shows that $(1 + q_2\a_2/\a_1) \to O(q_1)$
considering $c_1$. In fact, we find $\a_1\sim c_2/q_1^2$
 and $\a_2\sim -c_2/(q_1^2q_2)$, so $q_1^2\a_1$ and $q_2 \a_2/\a_1$ is finite.
 
At the colliding limit 
$|K_2' \rangle$ is reduced to $|I_2 \rangle$  in \eqref{G_2m} 
if the primary state $|\D \rangle$ has the conformal dimension
$\D_{012}$
which can be defined in terms of the $q$-state
\be
|\D_{012}   \rangle 
=\lim_{q_1, q_2 \to 0}q_1^{
-(\D_{012} -(\D_0+\D_1+\D_2)  +F_1(2)) } (1-q_2)^{-2\a_1\a_2}
| \vec a, \vec 0 ; \d_0(2)\,\rangle
\label{I2-primary}
\ee
where $\D_0+\D_1+\D_2- F_1(2)   = \d_0(2) + \Omega_0$
as in \eqref{F1(2)}.
We put the proper conformal dimension by multiplying 
the $q_1$ factor 
and remove the t-channel information by multiplying 
$(1-q_2)$ factor. 
The descendant  state $|\D_{012} +|\vec Y|  \rangle $ 
is obtained if one uses $| \vec a, \vec Y ; \d_0(2)\,\rangle$ in 
\eqref{I2-primary}. 
In addition, we can  use the freedom to put  normalization constant 
$(-\a_2/\a_1)^{ -(F_2(2)+2 \a_2\a_0) }$
so that $q_2$ factor  in front is 1 as $q_2 \to 0$.
Then the gauge conformal state we are preparing for the colliding limit is given as 
\be 
|K_2' \rangle=  
\sum_{\vec Y}  q_1^{|\vec Y|}\sqrt{\Z_{\text {vect}}(\vec a,\vec Y)}
 \bigg\{\sum_{\vec W}q_2^{|\vec W|}
\Z_{\text {vect}}(\vec b,\vec W)\Zbf(\vec a,\vec Y;\vec b,\vec W|\n)
\prod_{I=1,2} \Z_{\text {afd}}(\vec b, \vec W, \m_I) \bigg\} 
| \D_{012}+ |\vec Y| \rangle.
\label{K2-colliding}
\ee

Note that 
$c_2=\Lambda$ provides the overall scaling parameter. 
Therefore, we need $(q_1^2 \a_1)^{|\vec Y|/2}$ 
in the summation over $\vec Y$.
On the other hand,  for the summation over 
$\vec W$, we need $q_2$ dependent quantity. 
Note that there are  three other parameters 
$a_1, b_1$  and $t$  in \eqref{G_2m}.
All the quantities are to be given in
$c_0$, $c_1$ and $c_2$ which is finite at the colliding limit. 
In fact, $ a_1 \Lambda^{3/2}$ and $ t \Lambda$  
correspond to the eigenvalue
of $L_3$ and $L_2$, respectively.
$b_1$ is to be related with the normalization 
of the irregular state \cite{CRZ_2015}. 
The candidates of the $q_1$ independent terms
are  $c_0$  and  the  combination  
$c_1^2/c_2= \a_1 (1 +q_2 \a_2/\a_1)^2 $.
Therefore, $c_1^2/c_2$ is very tricky to get 
because  we need the combination 
of  $\a_1$ and $q_2 \a_2/\a_1$ 
at the colliding limit in the summation over $\vec W$ 
in \eqref{K2-colliding}. \footnote{If we send $b \to \n$ together with 
the colliding limit, we have
$
|T_2  ; \d_0 \rangle 
\sim \sum_{\vec Y, \vec W}  q_1^{|\vec Y|}a_1^{|\vec Y|}q_2^{|\vec W|}a_2^{|\vec W|}a_1^{-|\vec W|}
F   \,
|\vec a,\vec Y; \d_0(2) \rangle
$, with F finite. In this case the $\vec W$ related terms are finite.}

It is worth to note that at $ |\vec Y|=0$ 
the coefficient in \eqref{K2-colliding}
is not 1.
Therefore, 
 at the colliding limit one has the inner product
\be 
\langle \D_{012}  |K_2' \rangle=  
\sum_{\vec W}q_2^{|\vec W|}
\Z_{\text {vect}}(\vec b,\vec W)\Zbf(\vec a,\vec 0;\vec b,\vec W|\n)
\prod_{I=1,2} \Z_{\text {afd}}(\vec b, \vec W, \m_I) \langle \D_{012}  | \D_{012}  \rangle,
\ee
which is not a simple constant but 
should be related with the  partition function 
$Z_{(02)}$ of the irregular matrix model \cite{CRZ_2015, CR_2014}.

For the general state of rank $m$, 
one can start with the Liouville state 
\ba
|R'_m\rangle=\prod_{r=1}^{m}(z_r)^{-2\a_r\a_0}\times\prod_{i<j}^{m}(z_i-z_j)^{-2\a_i\a_j}|\,R_m(z)\,\rangle
\ea
so that $|R'_m\rangle$ satisfies 
\begin{align}
\cL_1| R'_m(z) \rangle
=& \sum_{r=1}^{m}\left(z_r^{2}
\frac{\partial}{\partial z_r}
+2z_r\a_r(\sum_{k=0}^m\a_k-Q)\right)
|R'_m(z) \rangle
\\ 
\cL_2| R'_m(z) \rangle
=& \sum_{r=1}^{m}\left(z_r^{3}
\frac{\partial}{\partial z_r}
+z_r^{2}\a_r(\a_r+2\sum_{k=0}^m\a_k-3Q)\right)
+2\sum_{i<j}^{m}z_iz_j{\a_i\a_j}|\,R'_m(z)\,\rangle .
\end{align}
At the colliding limit we have  
finite variables $ c_k=\sum_{i=1}^m z_i^k\a_i$ and 
\begin{align}
\cL_1|I_m(z) \rangle
&=\sum_{l=1}^{m-1}lc_{l+1}\frac{\partial}{\partial c_{l}}
+2c_1(Q-c_0)| I_m(z) \rangle
\\ 
\cL_2|\,I_m(z)\,\rangle
&=\sum_{l=1}^{m-2}lc_{l+2}
\frac{\partial}{\partial c_{l}}+(3Q-2c_0)c_2-c_1^2
|\,I_m(z) \rangle.
\end{align}

Similarly, we define the $q$-state 
\be
|K_m' \rangle
=q_1^{(\sum_{i=0}^m\D_i)-\D}f(q_{j\neq1})|K_m \rangle
\ee
with $|K_m \rangle$ defined in \eqref{TmRm}.
Explicitly, we have 
\begin{align}
\label{KmTm}
|K_m' \rangle&=  \prod_{r=1}^{m-1}
(q_1\cdots q_r)^{\frac{1}{{\e_1\e_2}}
\big({\frac12}\sum_p^2 (a^{(r)}_{p})^2
-{\frac12}\sum_p^2 (a^{(r+1)}_{p})^2+(3\m_{r,r+1}-5\e_+
+2\sum_{i=1}^{r-1}\m_{i,i+1})(\e_+-\m_{r,r+1})-2(\m_1-\m_2)(\e_+- \m_{r,r+1}) \big)}
\nn\\
&~~\times(q_1\cdots q_m)^{\frac{1}{{\e_1\e_2}}\big(\frac12\sum_p (a^{(m)}_{p})^2+(\m_1+\m_2)(-\e_++\sum_{i=1}^{m-1}\m_{i,i+1})-\m_1\m_2-2\m_1^2\big)}\\
&~~\times\prod_{i<j}^{m-1}(q_1\cdots q_i-q_1\cdots q_j)^{{\frac{1}{{\e_1\e_2}}\big(4(\m_{i,i+1}-\e_+)(\e_+-\m_{j,j+1}) \big)}}\times\prod_{r}^{m-1}(q_1\cdots q_i-q_1\cdots q_m)^{{\frac{1}{{\e_1\e_2}}\big(2(\m_{r,r+1}-\e_+)(\m_1+\m_2) \big)}}
|T_m\rangle.
\nn
\end{align}

As noted in rank 2, the holomorphic coordinates 
should have the hierarchical structure
$z_m<z_{m-1}< \cdots<z_0 \to 0$
at the colliding limit. 
This s-channel limit is obtained if 
all $q_i \to 0$. 
$q_1$ dependence takes care of the proper scaling and 
disappears. In addition, t-channel quantity
(powers of $(1-q_i)$) 
should be absorbed into the definition of the primary q-state
$| \D_{01\cdots m}\rangle$.  
Then we are left with $ |K_m' \rangle =|T_m\rangle$ 
where the factor  $\prod_{r=2}^{m-1} ( q_r)^{H_r}$
in \eqref{KmTm} 
is normalized as 1 by multiplying the appropriate constant.
In this case, the inner product  $\langle  \D_{01\cdots m}|K_m' \rangle$
at the colliding limit
is identified as 
the partition function $Z_{0m}$ of
 irregular matrix 
which is now given by summing Young diagrams. 

It is interesting to apply the Heisenberg algebra to 
$q$-state $|T_m\rangle$.
Using $\left[L_1, J_k\right] = -k J_{k+1}$ 
one  finds  
\ba
J_k|T_m\rangle&&=
\sum_{i=1}^{m-1}\frac{(q_1\cdots q_i)^k}{\sqrt{-\e_1\e_2}}\bigg\{N (\e_+ -\n_{i})\bigg\}+\frac{(q_1\cdots q_m)^k}{\sqrt{-\e_1\e_2}}\bigg\{ \sum_I\m_I \bigg\}|T_m\rangle\,.
\ea
At the colliding limit, one has 
$J_k|T_m\rangle =0 $ when $k >m$ but 
\be
J_k|T_m\rangle = c_k|T_m\rangle
\ee
for $1 \le k \le m$. 
Therefore, 
$|T_m\rangle$ becomes  the coherent state of
Heisenberg algebra  at the colliding limit.

\section{Conclusion}
We  construct gauge conformal state based on AFLT basis 
and interwiners for the spherical Hecke algebra with central extension. 
The $q$-coordinate is the instanton expansion parameter 
and  Hecke algebra has the $q$-differential representation 
on the $q$-state. 
The $q$-representation is used to find the exact relation 
with the Liouville conformal state
where conformal scaling is to be carefully matched. 
The $q$-state reduces to the irregular conformal state 
at the colliding limit,
which provides the formal structure of the irregular state
and its inner product is identified with the partition 
of the irregular matrix. 
However, it is not yet clear how to get the explicit 
summation over Young diagram. 

Our study has been limited to the Virasoro conformal state
which is related with SU(2) gauge group. 
This method can be extended to $\cW$ 
coformal state without any difficulty. 
The $q$-state  for SU(N) gauge group can be 
obtained by extending the Young diagrams. 
Using $D_{-r,s}$, one has the actions of $\cW^{(s)}_{r}$ operators 
on the SU(N) gauge conformal state. 
Besides, generalization to 5 dimensions seems natural, 
using the 5D version of SH \cite{BFMZZ}.

\subsection*{Acknowledgements}
The authors thank Y. Matsuo and J.-E. Bourgine for useful discussions.
This work is supported by the National Research Foundation of Korea(NRF) grant funded by the Korea government(MSIP) (NRF-2014R1A2A2A01004951).

\appendix
\section{Calculation details}
All the following holds for SU(N) case.
\subsection{The component for the instanton part of Nekrasov Partition function}
\begin{eqnarray}\label{Zbfd}
\Zbf(\vec{a},\vec Y;\vec{b},\vec{W}|m_{12})
&=& \prod_{\ell=1}^{N_1} \prod_{\ell'=1}^{N_2} g_{Y_\ell,W_{\ell'}}(a_\ell-b_{\ell'}-m_{12})\\
g_{\l,\mu}(x)&=& \prod_{(i,j)\in \l}(x+\e_1(\l_j'-i+1)-\e_2(\mu_i-j))\prod_{(i,j)\in \mu}(-x+\e_1(\mu_j'-i)-\e_2(\l_i-j+1))\,.
\end{eqnarray}
Here $\l_i$ is the height of $i^\mathrm{th}$ column and $\l'_i$ is the length of $i^\mathrm{th}$ row of Young diagram $\l$
\begin{eqnarray}\label{def_Zf}
\Z_\mathrm{fund}(\vec{m};\vec{a},\vec{Y})&=&\Zbf(\vec{m},\vec \emptyset;\vec{a},\vec{Y}|0).
\end{eqnarray}
\begin{eqnarray}\label{def_Zaf}
\Z_\mathrm{afd}(\vec{m};\vec{a},\vec{Y})&=&\Zbf(\vec{a},\vec{Y};-\vec{m},\vec \emptyset|0)=\Z_\mathrm{fund}(-\e_+-\vec m;\vec{a},\vec{Y}).
\end{eqnarray}
\begin{eqnarray}
\Zv(\vec{a},\vec{Y})&=& \Zbf(\vec{a},\vec{Y};\vec{a},\vec{Y}|0)^{-1}.
\end{eqnarray}
\subsection{Useful formulas}
We use the formulas in \cite{BMZ_2015} 
\begin{eqnarray}
\frac{\Zbf(\vec a, \vec Y+x;\vec b, \vec W|\n)}
{\Zbf(\vec a, \vec Y;\vec b, \vec W|\n)}&=&
\frac{\prod_{y\in A(\vec W)} 
(\phi_x-\phi_y+\e_+ -\n)}{\prod_{y\in R(\vec W)} (\phi_x-\phi_y-\n)},
\label{prop_Zbf1}\\
\frac{\Zbf(\vec a, \vec Y-x;\vec b, \vec W|\n)}
{\Zbf(\vec a, \vec Y;\vec b, \vec W|m)}&=&
\frac{\prod_{y\in R(\vec W)} 
(\phi_x-\phi_y-\n)}{\prod_{y\in A(\vec W)} (\phi_x-\phi_y+\e_+ -\n)},
\label{prop_Zbf2}\\
\frac{\Zbf(\vec a, \vec Y;\vec b, \vec W+x|\n)}
{\Zbf(\vec a, \vec Y;\vec b, \vec W|\n)}&=&
\frac{\prod_{y\in A(\vec Y)} 
(\phi_x-\phi_y+\n)}{\prod_{y\in R(\vec Y)} (\phi_x-\phi_y+\n-\e_+)},
\label{prop_Zbf3}\\
\frac{\Zbf(\vec a, \vec Y;\vec b, \vec W-x|\n)}
{\Zbf(\vec a, \vec Y;\vec b, \vec W|\n)}&=&
\frac{\prod_{y\in R(\vec Y)} 
(\phi_x-\phi_y+\n-\e_+)}{\prod_{y\in A(\vec Y)} (\phi_x-\phi_y+\n)},
\label{prop_Zbf4}
\end{eqnarray}
\begin{eqnarray}
\dfrac{\Zv(\vec a,\vec Y+x)}{\Zv(\vec a,\vec Y)}&=&-\dfrac1{\e_1\e_2}\dfrac{\prod_{y\in R(\vec Y)}(\phi_x-\phi_y)(\phi_x-\phi_y-\e_+)}{\prod_{\superp{y\in A(\vec Y)}{y\neq x}}(\phi_x-\phi_y)(\phi_x-\phi_y+\e_+)},\label{prop_Zv}\\
\quad \dfrac{\Zv(\vec a,\vec Y-x)}{\Zv(\vec a,\vec Y)}&=&-\dfrac1{\e_1\e_2}\dfrac{\prod_{y\in A(\vec Y)}(\phi_x-\phi_y)(\phi_x-\phi_y+\e_+)}{\prod_{\superp{y\in R(\vec Y)}{y\neq x}}(\phi_x-\phi_y)(\phi_x-\phi_y-\e_+)}\,.\label{prop_Zv2}
\end{eqnarray}
\begin{eqnarray}
\dfrac{Z_{\text {afd}}(\vec a, \vec Y+x, \m_I)}{Z_{\text {afd}}(\vec a, \vec Y, \m_I)}&=&(\phi_x+\m_I)
\end{eqnarray}

Also we know that for arbitratry $m$ and $
z_k= q_1\cdots q_k\,,\qquad(r=1, \cdots N)
$,
\ba
&& \bigg\{\sum_{k=1}^{N-1}(q_1\cdots q_k)^m(|\vec{Y}_k|-|\vec{Y}_{k+1}|)+(q_1\cdots q_N)^m|\vec{Y}_N|\bigg\}|T_m\rangle\\
&&=\bigg\{\sum_{k=1}^{N-1}(q_1\cdots q_k)^m(q_k\frac{\partial}{\partial q_k}-q_{k+1}\frac{\partial}{\partial q_{k+1}})+(q_1\cdots q_N)^m(q_N\frac{\partial}{\partial q_N})\bigg\}|T_m\rangle\nn\\
&&=\sum_{k=1}^{N}z_k^{m+1}\frac{\partial}{\partial z_k}|T_m\rangle\,.\nn
\ea
\subsection{Action on $| G, \vec a, \m_I\rangle$}
In the following we re-derive the rank 1 case using a method introduced in \cite{BMZ_2015}, as an alternative of \cite{MRZ_2014}. First notice that,
\ba
D_{-1,n}\,q_1^D| G, \vec a, \m_I\rangle=q_1^{D+1} D_{-1,n} | G, \vec a, \m_I\rangle\,.
\ea

Then from \eqref{action},
\ba
D_{-1,n}| G, \vec a, \m_I\rangle&=&\sum_{\vec Y}\sqrt{\Z_{\text {vect}}(\vec a,\vec Y)}\prod_{I=1}^N\Z_{\text {afd}}(\vec a, \vec Y, \m_I)  \sum_{x\in R(\vec{Y})}
(\phi_x)^n\L_x(\vec{Y})|\vec{a},\vec{Y}- x\rangle\\
&=&\sum_{\vec Y} \sum_{x\in A(\vec{Y})}\sqrt{Z_{\text {vect}}(\vec a,\vec Y+x)}\bigg(\prod_{I=1}^N Z_{\text {afd}}(\vec a, \vec Y+x, \m_I) \bigg)
(\phi_x)^n\L_x(\vec{Y}+x)|\vec{a},\vec{Y}\rangle \nn
\ea
using the fact that $\L_x(\vec Y+x)^2=\L_x(\vec Y)^2$, $\forall x\in A(\vec Y)$,
\ba
&&D_{-1,n}| G, \vec a, \m_I\rangle=\sum_{\vec Y} \sum_{x\in A(\vec{Y})}(\phi_x)^n\L_x(\vec{Y})\sqrt{\frac{Z_{\text {vect}}(\vec a,\vec Y+x)}{Z_{\text {vect}}(\vec a,\vec Y)}}\bigg(\prod_{I=1}^N\frac{Z_{\text {afd}}(\vec a, \vec Y+x, \m_I)}{Z_{\text {afd}}(\vec a, \vec Y, \m_I)}\bigg)\nn\\
&&\times
\sqrt{Z_{\text {vect}}(\vec a,\vec Y)}\bigg(\prod_{I=1}^NZ_{\text {afd}}(\vec a, \vec Y, \m_I) \bigg)\,|\vec a,\vec Y\rangle
\ea
After some calculation, we find
\ba
&&D_{-1,n}| G, \vec a, \m_I\rangle=\frac{1}{\sqrt{-\e_1\e_2}}\sum_{\vec Y} \sum_{x\in A(\vec{Y})}(\phi_x)^n\dfrac{\prod_{{w\in R(\vec{Y})}}(\phi_x-\phi_w-\e_+)}{\prod_{\superp{y\in A(\vec{Y})}{y\neq x}}(\phi_x-\phi_y)}\prod_{I=1}^N(\phi_x+\m_I)\nn\\
&&\times
\sqrt{Z_{\text {vect}}(\vec a,\vec Y)}\prod_{I=1}^NZ_{\text {afd}}(\vec a, \vec Y, \m_I)  \,|\vec a,\vec Y\rangle
\ea
By setting $x_I=\{\phi_y, (y \in A(Y^{(l)}) \}$ and $y_J=\{\phi_w+\e_+, (w \in R(Y^{(l)});-\m_1,\dots, -\m_N \}$, the above can
be simplified using the KMZ equation \cite{KMZ2012},
\ba
\label{math}
\sum_{I=1}^{\mathcal{N}} (x_I)^m \frac{\prod_{J=1}^{\mathcal{M}} (x_I-y_J)}{\prod_{J (\neq I)}^{\mathcal{N}} (x_I-x_J)}
=\sum_{n=0}^{m+1+{\mathcal{M}}-{\mathcal{N}}} f_{m-n+1+{\mathcal{M}}-{\mathcal{N}}}(-y) b_{n}(x) \,,
\ea
where
$
f_n(x)=\sum_{I_1<\cdots<I_n} x_{I_1}\cdots x_{I_n}
$ , 
and
$
b_n(x)=\sum_{I_1\leq\cdots \leq I_n }x_{I_1}\cdots x_{I_n}\,.$ 
In details, since
\ba
\big(\sum_I x_I- \sum_J y_J\big)=\sum_p^N a_p +\sum_{I=1}^N\m_I\,,
\ea
and
\ba
\frac12\big(\sum_{I} x_I^2-\sum_{J} y_J^2\big)&&=-\frac12\sum_p\bigg[ \sum_{k=1}^{f_p}\big( 2a_p+\e_1(r_k+r_{k-1})+2\e_2s_k)\big)(\e_1r_k-\e_1r_{k-1})+\frac12(a_p+\e_1r_f)^2\bigg]
-\frac12\sum_{I=1}^N\m_I^2\nn\\
&&=-\e_1\e_2|\vec{Y}|+\frac12\sum_p^N (a_p)^2-\frac12\sum_{I=1}^N\m_I^2\,,
\ea
we find
\ba
\sum_{n=0}^{2} f_{2-n}(-y) b_{n}(x) &&=\frac12\big(\sum_I x_I- \sum_J y_J\big)^2+\frac12\big(\sum_{I} x_I^2-\sum_{J} y_J^2\big)\\
&&=-\e_1\e_2|\vec{Y}|+\frac12\sum_p^N (a_p)^2-\frac12\sum_{I=1}^N\m_I^2+\frac12(\sum_p^N a_p +\sum_{I=1}^N\m_I)^2,\nn
\ea
which leads to \eqref{rank101} and \eqref{rank111} for SU(2) case.

Further for $D_{-2,d}$,  we know
\ba
D_{-2,0} &=& \left[D_{-1,0}\, , \,D_{-1,1} \right],\\
D_{-2,1} &=& \left[D_{-1,0} \, , \, D_{-1,2} \right],\\
D_{-2,d} &=& \left[D_{-1,0}\, , \,D_{-1,d+1} \right]-
\left[D_{-1,1} \, , \, D_{-1,d} \right] \,.
\ea
It reads that
\ba
&&D_{-1,m}D_{-1,n}| G, \vec a, \m_I\rangle\nn\\
&&=\sum_{\vec Y} \sum_{x\in A(\vec{Y})}\sqrt{Z_{\text {vect}}(\vec a,\vec Y+x)}\prod_{I=1}^N Z_{\text {afd}}(\vec a, \vec Y+x, \m_I)
(\phi_x)^n\L_x(\vec{Y})\sum_{t\in  R(\vec{Y})}
(\phi_t)^m\L_t(\vec{Y})|\vec{a},\vec{Y}- t\rangle\\
&&=\sum_{\vec Y}\sum_{x\in A(\vec{Y}+t)}\sqrt{Z_{\text {vect}}(\vec a,\vec Y+t+x)}\prod_{I=1}^N Z_{\text {afd}}(\vec a, \vec Y+t+x, \m_I)
(\phi_x)^n\L_x(\vec{Y}+t)\sum_{t\in  A(\vec{Y})}
(\phi_t)^m\L_t(\vec{Y})|\vec{a},\vec{Y}\rangle,\nn
\ea
Combining the above, we find
\ba\label{D20}
D_{-2,0}| G, \vec a, \m_I\rangle
&&=\sum_{\vec Y}\sum_{t\in  A(\vec{Y})}\sum_{x\in A(\vec{Y}+t)}\sqrt{Z_{\text {vect}}(\vec a,\vec Y+t+x)}\prod_{I=1}^N Z_{\text {afd}}(\vec a, \vec Y+t+x, \m_I)
\L_x(\vec{Y}+t)\L_t(\vec{Y})\nn\\
&&\times
\bigg(\phi_t-\phi_x\bigg)|\vec{a},\vec{Y}\rangle,
\ea
and for $n\geq 1$,
\ba
D_{-2,n}| G, \vec a, \m_I\rangle
&&=\sum_{\vec Y}\sum_{t\in  A(\vec{Y})}\sum_{x\in A(\vec{Y}+t)}\sqrt{Z_{\text {vect}}(\vec a,\vec Y+t+x)}\prod_{I=1}^N Z_{\text {afd}}(\vec a, \vec Y+t+x, \m_I)
\L_x(\vec{Y}+t)\L_t(\vec{Y})\nn\\
&&\times
\bigg((\phi_t)^n+(\phi_x)^n\bigg)\bigg(\phi_t-\phi_x\bigg)|\vec{a},\vec{Y}\rangle.
\ea
The key is to solve the following,
\ba
&& \sum_{x\in A(\vec{Y}+t)}(\phi_x)^n\L_x(\vec{Y}+t)\sqrt{\frac{Z_{\text {vect}}(\vec a,\vec Y+t+x)}{Z_{\text {vect}}(\vec a,\vec Y+t)}}\prod_{I=1}^N \frac{Z_{\text {afd}}(\vec a, \vec Y+t+x, \m_I)}{Z_{\text {afd}}(\vec a, \vec Y+t, \m_I)}\times
(\phi_x-\phi_t)\\
&&=\frac{1}{\sqrt{-\e_1\e_2}}\sum_{x\in A(\vec{Y}+t)}(\phi_x)^n\dfrac{\prod_{{w\in R(\vec{Y}+t)}}(\phi_x-\phi_w-\e_+)}{\prod_{\superp{y\in A(\vec{Y}+t)}{y\neq x}}(\phi_x-\phi_y)}\prod_{I=1}^N (\phi_x+\m_I)\times
(\phi_x-\phi_t)\nn\\
&&=\frac{1}{\sqrt{-\e_1\e_2}}\sum_{x\in A(\vec{Y}),x\neq t}(\phi_x)^n\dfrac{\prod_{{w\in R(\vec{Y})}}(\phi_x-\phi_w-\e_+)}{\prod_{\superp{y\in A(\vec{Y})}{y\neq x}}(\phi_x-\phi_y)}
\frac{(\phi_x-\phi_t-\e_+)(\phi_x-\phi_t)^2}{(\phi_x-\phi_t-\e_1)(\phi_x-\phi_t-\e_2)}
\prod_{I=1}^N (\phi_x+\m_i)
\nn\\
&&+\frac{1}{\sqrt{-\e_1\e_2}}(\phi_t+\e_2)^n\dfrac{\prod_{{w\in R(\vec{Y})}}(\phi_t+\e_2-\phi_w-\e_+)}{\prod_{\superp{y\in A(\vec{Y})}{y\neq t}}(\phi_t+\e_2-\phi_y)}
\frac{(-\e_1\e_2)}{(\e_2-\e_1)}
\prod_{I=1}^N (\phi_t+\e_2+\m_I)\nn\\
&&+\frac{1}{\sqrt{-\e_1\e_2}}(\phi_t+\e_1)^n\dfrac{\prod_{{w\in R(\vec{Y})}}(\phi_t+\e_1-\phi_w-\e_+)}{\prod_{\superp{y\in A(\vec{Y})}{y\neq t}}(\phi_t+\e_1-\phi_y)}
\frac{(-\e_1\e_2)}{(\e_1-\e_2)}
\prod_{I=1}^N (\phi_t+\e_1+\m_I)\,.\nn
\ea
By exchanging $x$ and $t$,
\ba
&& \sum_{t\in A(\vec{Y})}\dfrac{\prod_{{w\in R(\vec{Y})}}(\phi_t-\phi_w-\e_+)}{\prod_{\superp{y\in A(\vec{Y})}{y\neq t}}(\phi_t-\phi_y)}\sum_{x\in A(\vec{Y}),x\neq t}\dfrac{\prod_{{w\in R(\vec{Y})}}(\phi_x-\phi_w-\e_+)}{\prod_{\superp{y\in A(\vec{Y})}{y\neq x}}(\phi_x-\phi_y)}
\frac{(\phi_x-\phi_t-\e_+)(\phi_x-\phi_t)^2}{(\phi_x-\phi_t-\e_1)(\phi_x-\phi_t-\e_2)}
\prod_{I=1}^N (\phi_x+\m_I)\nn\\
&&=\sum_{t, x\in A(\vec{Y})}\dfrac{\prod_{{w\in R(\vec{Y})}}(\phi_t-\phi_w-\e_+)}{\prod_{\superp{y\in A(\vec{Y})}{y\neq t}}(\phi_t-\phi_y)}\dfrac{\prod_{{w\in R(\vec{Y})}}(\phi_x-\phi_w-\e_+)}{\prod_{\superp{y\in A(\vec{Y})}{y\neq x}}(\phi_x-\phi_y)}
\frac{(\phi_x-\phi_t-\e_+)(\phi_x-\phi_t)^2}{(\phi_x-\phi_t-\e_1)(\phi_x-\phi_t-\e_2)}
\prod_{I=1}^N (\phi_x+\m_I)\\
&&=-\sum_{t\in A(\vec{Y})}\dfrac{\prod_{{w\in R(\vec{Y})}}(\phi_t-\phi_w-\e_+)}{\prod_{\superp{y\in A(\vec{Y})}{y\neq t}}(\phi_t-\phi_y)}\sum_{x\in A(\vec{Y}),x\neq t}\dfrac{\prod_{{w\in R(\vec{Y})}}(\phi_x-\phi_w-\e_+)}{\prod_{\superp{y\in A(\vec{Y})}{y\neq x}}(\phi_x-\phi_y)}
\frac{(\phi_x-\phi_t+\e_+)(\phi_x-\phi_t)^2}{(\phi_x-\phi_t+\e_1)(\phi_x-\phi_t+\e_2)}
\prod_{I=1}^N (\phi_t+\m_I)\nn.
\ea
As a result\footnote{For $n=0$ there is an extra factor of $1/2$ due to \eqref{D20}.},{\small
\ba
&&D_{-2,n}| G, \vec a, \m_I\rangle = \nn\\
&& \bigg\{\sum_{t\in A(\vec{Y})}\frac{1}{2(\e_2-\e_1)}\big((\phi_t)^n+(\phi_t+\e_2)^n\big)\dfrac{\prod_{{w\in R(\vec{Y})}}(\phi_t-\phi_w-\e_+)}{\prod_{\superp{y\in A(\vec{Y})}{y\neq t}}(\phi_t-\phi_y)}\dfrac{\prod_{{w\in R(\vec{Y})}}(\phi_t+\e_2-\phi_w-\e_+)}{\prod_{\superp{y\in A(\vec{Y})}{y\neq t}}(\phi_t+\e_2-\phi_y)}
\prod_{I=1}^N(\phi_t+\m_I)(\phi_t+\e_2+\m_I)\nn\\
&&+\sum_{t\in A(\vec{Y})}\frac{1}{2(\e_1-\e_2)}\big((\phi_t)^n+(\phi_t+\e_1)^n\big)\dfrac{\prod_{{w\in R(\vec{Y})}}(\phi_t-\phi_w-\e_+)}{\prod_{\superp{y\in A(\vec{Y})}{y\neq t}}(\phi_t-\phi_y)}\dfrac{\prod_{{w\in R(\vec{Y})}}(\phi_t+\e_1-\phi_w-\e_+)}{\prod_{\superp{y\in A(\vec{Y})}{y\neq t}}(\phi_t+\e_1-\phi_y)}
\prod_{I=1}^N(\phi_t+\m_I)(\phi_t+\e_1+\m_I)\nn\\
&& -\sum_{t\in A(\vec{Y})}\frac{1}{2(\e_2-\e_1)}\big((\phi_t)^n+(\phi_t-\e_2)^n\big)\dfrac{\prod_{{w\in R(\vec{Y})}}(\phi_t-\phi_w-\e_+)}{\prod_{\superp{y\in A(\vec{Y})}{y\neq t}}(\phi_t-\phi_y)}\dfrac{\prod_{{w\in R(\vec{Y})}}(\phi_t-\e_2-\phi_w-\e_+)}{\prod_{\superp{y\in A(\vec{Y})}{y\neq t}}(\phi_t-\e_2-\phi_y)}
\prod_{I=1}^N(\phi_t+\m_I)(\phi_t-\e_2+\m_I)\nn\\
&&-\sum_{t\in A(\vec{Y})}\frac{1}{2(\e_1-\e_2)}\big((\phi_t)^n+(\phi_t-\e_1)^n\big)\dfrac{\prod_{{w\in R(\vec{Y})}}(\phi_t-\phi_w-\e_+)}{\prod_{\superp{y\in A(\vec{Y})}{y\neq t}}(\phi_t-\phi_y)}\dfrac{\prod_{{w\in R(\vec{Y})}}(\phi_t-\e_1-\phi_w-\e_+)}{\prod_{\superp{y\in A(\vec{Y})}{y\neq t}}(\phi_t-\e_1-\phi_y)}
\prod_{I=1}^N(\phi_t+\m_I)(\phi_t-\e_1+\m_I)\bigg\}| G, \vec a, \m_I\rangle\nn\\
&&=\frac{1}{2(\e_1-\e_2)}\bigg\{\e_1\sum_{i=1}^2\sum_{I=1}^{\mathcal{N}} (x^{(i)}_I)^n \frac{\prod_{J=1}^{\mathcal{M}} (x^{(i)}_I-y^{(i)}_J)}{\prod_{J (\neq I)}^{\mathcal{N}} (x^{(i)}_I-x^{(i)}_J)}-\e_2\sum_{i=3}^4\sum_{I=1}^{\mathcal{N}} (x^{(i)}_I)^n \frac{\prod_{J=1}^{\mathcal{M}} (x^{(i)}_I-y^{(i)}_J)}{\prod_{J (\neq I)}^{\mathcal{N}} (x^{(i)}_I-x^{(i)}_J)}\bigg\}| G, \vec a, \m_I\rangle\,,
\ea}
with $x_I=\{\phi_y, (y \in A(Y^{(l)}) \}$ and $y_J=\{\phi_w+\e_+, (w \in R(Y^{(l)});-\m_1,\dots, -\m_N \}$,\\
$x^{(1)}_I=\{x_I; x_I-\e_1\}$ and $y^{(1)}_J=\{y_I; y_I-\e_1\}$, 
$x^{(2)}_I=\{x_I; x_I+\e_1\}$ and $y^{(2)}_J=\{y_I; y_I+\e_1\}$,\\
$x^{(3)}_I=\{x_I; x_I-\e_2\}$ and $y^{(3)}_J=\{y_I; y_I-\e_2\}$, 
$x^{(4)}_I=\{x_I; x_I+\e_2\}$ and $y^{(4)}_J=\{y_I; y_I+\e_2\}$.\\
Especially, 
\ba
D_{-2,0}| G, \vec a, \m_i\rangle =\big(\sum_I x_I- \sum_J y_J\big),
\ea
and
\ba
D_{-2,1}|| G, \vec a, \m_i\rangle =2\big(\sum_I x_I- \sum_J y_J\big)^2+\big(\sum_{I} x_I^2-\sum_{J} y_J^2\big).
\ea
Thus we obtain \eqref{rank201} and \eqref{rank211}.
\subsection{Action on $| T_2\rangle$}
The proofs are given below. By definition,
\ba
D_{-1,n}|T_2\rangle&=&D_{-1,n}\,q_1^DV_{12}\,q_2^D| G, \vec a, \m_I\rangle=q_1^{D+1}[D_{-1,n},V_{12}]q_2^D| G, \vec a, \m_I\rangle+q_1^{D+1}V_{12}q_2^{D+1}D_{-1,n}| G, \vec a, \m_I\rangle\,.
\ea
Since
\begin{equation}
D_{-1,n}V(\vec a,\vec b|\n)=\sum_{\vec Y,\vec X}\sum_{x\in A(\vec Y)}(\phi_x)^n\L_x(\vec Y)
\frac{\bZbf(\vec a,\vec Y+x;\vec b,\vec X|\n)}{\bZbf(\vec a,\vec Y;\vec b,\vec X|\n)} 
\bZbf(\vec a,\vec Y;\vec b,\vec X|\n)
|\vec a,\vec Y\rangle\langle \vec b+\n,\vec X|\,,
\end{equation}
with
\begin{equation}
\L_x(\vec Y)
\frac{\bZbf(\vec a,\vec Y+x;\vec b,\vec  X|\n)}{\bZbf(\vec a,\vec Y;\vec b,\vec  X|\n)}
=\frac{1}{\sqrt{-\e_1\e_2}}\frac{\prod_{y\in R(\vec Y)}(\phi_x-\phi_y-\e_+)}{\prod_{\superp{y\in A(\vec Y)}{y\neq x}}(\phi_x-\phi_y)}
\frac{\prod_{y\in A(\vec X)}(\phi_x-\phi_y-\n+\e_+)}{\prod_{y\in R(\vec X)}(\phi_x-\phi_y-\n)}\,.
\end{equation}
And
\begin{equation}
-V(\vec a,\vec b|\n)D_{-1,n}=\sum_{\vec Y,\vec X}\sum_{x\in R(\vec X)}(\phi_x+\n)^n{\L_x(\vec X)}
\frac{\bZbf(\vec a,\vec Y;\vec b,\vec X-x|\n)}{\bZbf(\vec a,\vec Y;\vec b,\vec  X|\n)} 
\bZbf(\vec a,\vec Y;\vec b,\vec  X|\n)
|\vec a,\vec Y\rangle\langle \vec b+\n,\vec X|,
\end{equation}
with
\begin{equation}
\L_x(\vec X)
\frac{\bZbf(\vec a,\vec Y;\vec b,\vec X-x|\n)}{\bZbf(\vec a,\vec Y;\vec b,\vec X|\n)}
=\frac{1}{\sqrt{-\e_1\e_2}}\frac{\prod_{y\in A(\vec X)}(\phi_x-\phi_y+\e_+)}{\prod_{\superp{y\in R(\vec X)}{y\neq x}}(\phi_x-\phi_y)}
\frac{\prod_{y\in R(\vec Y)}(\phi_x-\phi_y+\n-\e_+)}{\prod_{y\in A(\vec Y)}(\phi_x-\phi_y+\n)}\,.
\end{equation}
So we find
\begin{eqnarray}
&&[D_{-1,n},V_{12}]=\sum_{x\in A(\vec Y)}(\phi_x)^n\frac{\prod_{y\in R(\vec Y)}(\phi_x-\phi_y-\e_+)}{\prod_{\superp{y\in A(\vec Y)}{y\neq x}}(\phi_x-\phi_y)}
\frac{\prod_{y\in A(\vec X)}(\phi_x-\phi_y-\n+\e_+)}{\prod_{y\in R(\vec X)}(\phi_x-\phi_y-\n)}
\nonumber\\
&&+\sum_{x\in R(\vec X)}(\phi_x+\n)^n
\frac{\prod_{y\in A(\vec X)}(\phi_x-\phi_y+\e_+)}{\prod_{\superp{y\in R(\vec X)}{y\neq x}}(\phi_x-\phi_y)}
\frac{\prod_{y\in R(\vec Y)}(\phi_x-\phi_y+\n-\e_+)}{\prod_{y\in A(\vec Y)}(\phi_x-\phi_y+\n)}\nonumber\\
&&~~~~~=\sum_{I=1}^{\mathcal{N}} (x_I)^n \frac{\prod_{J=1}^{\mathcal{M}} (x_I-y_J)}{\prod_{J (\neq I)}^{\mathcal{N}} (x_I-x_J)}
\end{eqnarray}
with $\tilde x_I=\{\phi_y, (y \in A(Y^{(l)}); \phi_w+\n, (w \in R(X^{(l)}) \}$ and $\tilde y_J=\{\phi_w+\e_+, (w \in R(Y^{(l)});\phi_y-\e_++\n, (y \in A(X^{(l)}) \}$. Explicitly,
\ba
\big(\sum_I \tilde x_I- \sum_J \tilde y_J\big)=\sum_k^N (a_k-b_k+\e_+ -\n)\,,
\ea
and
\ba
\frac12\big(\sum_{I} \tilde x_I^2-\sum_{J} \tilde y_J^2\big)=-\e_1\e_2(|\vec{Y}|-|\vec{X}|)+\frac12\sum_p^N (a_p)^2-{\frac12}\sum_p^N (b_p-\e_+ +\n)^2\,,
\ea
thus
\begin{eqnarray}
&&[D_{-1,1},V_{12}]=\sum_{n=0}^{2} f_{2-n}(-y) b_{n}(x)\nonumber\\
&&=-\e_1\e_2(|\vec{Y}|-|\vec{X}|)+{\frac12}\sum_p^N\bigg[a_p^2-(b_p-\e_+ +\n)^2\bigg]+{\frac12}\bigg(\sum_p^N (a_p-b_p+\e_+ -\n)\bigg)^2 .
\end{eqnarray}
So we find \eqref{rank112}.
Similarly,
\ba
D_{-2,n}|T_2\rangle&=&q_1^{D+2}[D_{-2,n},V_{12}]q_2^D| G, \vec a, \m_I\rangle+q_1^{D+2}V_{12}q_2^{D+2}D_{-2,n}| G, \vec a, \m_I\rangle\,.
\ea
Here
\ba
&&[D_{-2,n},V_{12}] \nn\\
&&=\frac{1}{2(\e_1-\e_2)}\bigg\{\e_1\sum_{i=1}^2\sum_{I=1}^{\mathcal{N}} (\tilde x^{(i)}_I)^n \frac{\prod_{J=1}^{\mathcal{M}} (\tilde x^{(i)}_I-\tilde y^{(i)}_J)}{\prod_{J (\neq I)}^{\mathcal{N}} (\tilde x^{(i)}_I-\tilde x^{(i)}_J)}-\e_2\sum_{i=3}^4\sum_{I=1}^{\mathcal{N}} (\tilde x^{(i)}_I)^n \frac{\prod_{J=1}^{\mathcal{M}} (\tilde x^{(i)}_I-\tilde y^{(i)}_J)}{\prod_{J (\neq I)}^{\mathcal{N}} (\tilde x^{(i)}_I-\tilde x^{(i)}_J)}\bigg\}\,,
\ea
with
$\tilde x^{(1)}_I=\{\tilde x_I; \tilde x_I-\e_1\}$ and $\tilde y^{(1)}_J=\{\tilde y_I; \tilde y_I-\e_1\}$,
$\tilde x^{(2)}_I=\{\tilde x_I; \tilde x_I+\e_1\}$ and $y^{(2)}_J=\{\tilde y_I; \tilde y_I+\e_1\}$,\\
$\tilde x^{(3)}_I=\{\tilde x_I; \tilde x_I-\e_2\}$ and $y^{(3)}_J=\{\tilde y_I; \tilde y_I-\e_2\}$,
$\tilde x^{(4)}_I=\{\tilde x_I; \tilde x_I+\e_2\}$ and $y^{(4)}_J=\{\tilde y_I; \tilde y_I+\e_2\}$.\\
Then
\ba
&&[D_{-2,1},V_{12}] =2\big(\sum_I \tilde x_I- \sum_J \tilde y_J\big)^2+\big(\sum_{I} \tilde x_I^2-\sum_{J} \tilde y_J^2\big)\nn\\
&&=-2\e_1\e_2(|\vec{Y}|-|\vec{X}|)+\sum_p^N\bigg[a_p^2-(b_p-\e_+ +\n)^2\bigg]+2\bigg(\sum_p^N (a_p-b_p+\e_+ -\n)\bigg)^2.
\ea
This leads to \eqref{rank212}.


\begin{thebibliography}{99}
\bibitem{AGT_2009}
  L.~F.~Alday, D.~Gaiotto and Y.~Tachikawa,
 \emph{Liouville Correlation Functions from Four-dimensional Gauge Theories},
Lett.\ Math.\ Phys. {\bf 91} (2010) 167 [arXiv:0906.3219 [hep-th]].

\bibitem{Wyllard_2009}
N. Wyllard,  \emph{$A_{N-1} $ conformal Toda field theory correlation functions from conformal N = 2
SU(N) quiver gauge theories}, JHEP {\bf 11}  (2009) 002 [arXiv:0907.2189].

\bibitem{Gaiotto_2009} 
D. Gaiotto, \emph{Asymptotically free N=2 theories and irregular conformal blocks,} 
J.\ Phys.\ Conf.\ Ser.\  {\bf 462}  (2013)  012014
[arXiv:0908.0307 [hep-th]].

\bibitem{AD}
P. C. Argyres and M. R. Douglas, 
\emph{New phenomena in SU(3) supersymmetric gauge theory},
{Nucl. Phys. B} {\bf 448} (1995) 93  [arXiv:9505062 [hep-th]].

\bibitem{APSW}
 P. C. Argyres, M. R. Plesser, N. Seiberg and E. Witten, \emph{New N=2 superconformal field
theories in four-dimensions}, {Nucl. Phys. B} {\bf  461} (1996) 71,  arXiv: 9511154[hep-th].

\bibitem{Whittaker}
E. Felinska, Z. Jaskolski, and M. Kosztolowicz, \emph{ Whittaker Pairs for the Virasoro Algebra and 
the Gaiotto - Bmt States, } J. Math. Phys. {\bf 53} (2012) 033504 [arXiv:1112.4453 [math-ph]].


\bibitem{MMM2009}
A. Marshakov, A. Mironov and  A. Morozov, 
\emph{ On non-conformal limit of the AGT relations},
 Phys. Lett. B {\bf 682} (2009) 125-129  
[arXiv:0909.2052].


\bibitem{BMT_2011}
G. Bonelli, K. Maruyoshi and A. Tanzini,
\emph{Wild Quiver Gauge Theories}, JHEP {\bf 02}  (2012) 031 
[arXiv:1112.1691].

\bibitem{KMST2013}
H. Kanno, K. Maruyoshi, S. Shiba, and M. Taki, {\it W3 irregular states and isolated N=2
superconformal field theories}, JHEP {\bf 03} (2013) 147 [arXiv:1301.0721 [hep-th]].


\bibitem{GT2012}
D. Gaiotto and J. Techner, \emph{Irregular singularities in Liouville theory and Argyres-Douglas type gauge theories, I}, JHEP {\bf 12} (2012) 050 
[arXiv:1203.1052 [hep-th]].

\bibitem{EM_2009} 
T. \  Eguchi and  K.\  Maruyoshi,
\emph{Penner Type Matrix Model and Seiberg-Witten Theory},
 JHEP  {\bf 02}  (2010) 022 [arXiv:0911.4797 [hep-th]].

\bibitem{CRZ_2015}
S.-K.\ Choi, C.\  Rim and H.\  Zhang,
\emph{Virasoro irregular conformal block 
and beta deformed random matrix model}, 
Phys.\ Lett.\  B {\bf 742} (2015) 50 [arXiv:1411.4453 [hep-th]]. 

\bibitem{CRZ_2016}
S.-K.\ Choi and C.\  Rim,
\emph{Irregular matrix model with W symmetry },
J.\ Phys.\ A {\bf  49} (2016) 075201 
[arXiv:1506.03561 [hep-th]]; 
S.-K.\ Choi, C.\  Rim and H.\  Zhang,
\emph{Irregular conformal block, spectral curve and flow equations},
JHEP  {\bf03} (2016) 118  [arXiv:1510.09060 [hep-th]].

\bibitem{r:SV} O. Schiffmann and E. Vasserot, 
\emph{Cherednik algebras, W algebras and the equivariant cohomology of the moduli space of instantons on $A^2$}, [arXiv:1202.2756]. 

\bibitem{MRZ_2014}
Y.~Matsuo, C.~Rim, and H.~Zhang, \emph{Construction of Gaiotto states with
  fundamental multiplets through Degenerate DAHA},
 {{ JHEP} {\bf 09}
  (2014) 028}  [arXiv:1405.3141  [hep-th]].

\bibitem{Alba2011}
V.~A. Alba, V.~A. Fateev, A.~V. Litvinov, and G.~M. Tarnopolskiy, ``{On
  combinatorial expansion of the conformal blocks arising from AGT
  conjecture},'' {{  Lett.Math.Phys.} {\bf 98} (2011) 33--64},
[arXiv:1012.1312 [hep-th]];  
  V.~A.~Fateev and A.~V.~Litvinov,
  ``Integrable structure, W-symmetry and AGT relation,''
  JHEP {\bf 01} (2012) 051 
  [arXiv:1109.4042 [hep-th]].

\bibitem{MO_2012} 
D. Maulik and A. Okounkov, \emph{Quantum Groups and Quantum Cohomology}, arXiv:1211.1287
[math.AG].

\bibitem{BMZ_2015} 
J.-E. Bourgine, Y.~Matsuo and H.~Zhang,
\emph{Holomorphic field realization of SH$^c$ and quantum geometry of quiver gauge theories}, JHEP {\bf 04} (2016) 167
[arXiv:1512.02492 [hep-th] ]. 




\bibitem{Kanno:2013aha} 
  S.~Kanno, Y.~Matsuo and H.~Zhang,
 \emph{Extended Conformal Symmetry and Recursion Formulae for Nekrasov Partition Function},
  JHEP {\bf 08}, (2013) 028 
  [arXiv:1306.1523 [hep-th]].


\bibitem{KMZ2012} 
  S.~Kanno, Y.~Matsuo and H.~Zhang,
   \emph{Virasoro constraint for Nekrasov instanton partition function,}
  JHEP {\bf 10}, (2012) 097 
  [arXiv:1207.5658 [hep-th]].





\bibitem{Bourgine2014a}
J.-E. Bourgine, \emph{Spherical Hecke algebra in the Nekrasov-Shatashvili
  limit}, {{ JHEP}
  {\bf 01} (2015) 114}, [  arXiv:1407.8341 [hep-th]].


\bibitem{CR_2014}
S.-K Choi and C. Rim,
\emph{Parametric dependence of irregular conformal block},
JHEP {\bf 04} (2014) 106 [arXiv:1312.5535 [hep-th]].

\bibitem{BFMZZ} 
J.-E. Bourgine, M, Fukuda, Y.~Matsuo, H.~Zhang and R-D, Zhu,
\emph{Coherent states in quantum $W_{1+\infty}$  algebra and qq-character for 5d Super Yang-Mills}, 
[arXiv:1606.08020 [hep-th]]. 




\end{thebibliography}
\end{document}